\documentclass[preprintnumbers, aip,pop,superscriptaddress,amsmath,amssymb,reprint]{revtex4-1}

\usepackage{graphicx}
\usepackage{subfigure}  

\begin{document}

\preprint{LA-UR-16-29528}

\title{Observation and modeling of interspecies ion separation in inertial confinement fusion implosions via imaging x-ray spectroscopy}


\author{T. R. Joshi}
\email[]{tjoshi@lanl.gov}
\author{P. Hakel}
\author{S. C. Hsu}
\email[]{scotthsu@lanl.gov}
\author{E. L. Vold}
\author{M. J. Schmitt}
\author{N. M. Hoffman}
\author{R. M. Rauenzahn}
\affiliation{Los Alamos National Laboratory, Los Alamos, New Mexico 87545, USA}
\author{\\G. Kagan}
\author{X.-Z. Tang}
\affiliation{Los Alamos National Laboratory, Los Alamos, New Mexico 87545, USA}
\author{R. C. Mancini}
\affiliation{Physics Department, University of Nevada, Reno, Nevada 89557, USA}
\author{Y. Kim}
\author{H. W. Herrmann}
\affiliation{Los Alamos National Laboratory, Los Alamos, New Mexico 87545, USA}


\date{\today}

\begin{abstract}
We report first direct experimental evidence of interspecies ion separation in direct-drive ICF experiments performed at the OMEGA laser facility via spectrally, temporally and spatially resolved imaging x-ray-spectroscopy data [S. C. Hsu $et$ $al.$, EPL {\bf 115}, 65001 (2016)]. These experiments were designed based on the expectation that interspecies ion thermo-diffusion would be strongest for species with large mass and charge difference. The targets were spherical plastic shells filled with D$_2$ and a trace amount of Ar (0.1${\%}$ or 1${\%}$ by atom). Ar K-shell spectral features were observed primarily between the time of first-shock convergence and slightly before neutron bang time, using a time- and space-integrated spectrometer, a streaked crystal spectrometer, and two gated multi-monochromatic x-ray imagers fielded along quasi-orthogonal lines of sight. Detailed spectroscopic analyses of spatially resolved Ar K-shell lines reveal deviation from the initial 1{$\%$} Ar gas fill and show both Ar-concentration enhancement and depletion at different times and radial positions of the implosion. The experimental results are interpreted with radiation-hydrodynamic simulations that include recently implemented, first-principles models of interspecies ion diffusion. The experimentally inferred Ar-atom fraction profiles agree reasonably with calculated profiles associated with the incoming and rebounding first shock. \end{abstract}


\maketitle


\section{Introduction}
In inertial confinement fusion (ICF) experiments, nuclear fusion reactions are initiated by the heating  and  compressing  of  a  fuel  target,  typically  in  the  form  of  a  spherical plastic (or glass) shell filled with a combination of lower-$Z$ elements such as deuterium-tritium (DT), deuterium-helium 3 (D$ ^{3}$He) as fusion fuels.\cite{lindl_pop_04, s_atzeni_ppfc_09} Sometimes a little amount of higher-$Z$ elements like argon (Ar) or krypton (Kr) have also been used in the core along with the main fuel materials for diagnostic purposes because of their proven usefulness in x-ray spectroscopy.\cite{Golovkin02prl, Nagayama11jap, Nagayama14pop, Regan02pop, Welser-Sherrill07pre, Florido11pre, Haynes96pre} The target surface is irradiated either directly by laser beams or indirectly by x-rays obtained from the inner surface of a hohlraum, which, in turn, ablates the outer surface material of the target and implodes the remaining shell and fuel due to the rocket-like blow-off of the hot surface material. This process is designed to create several shocks that travel inward across the core. The implosion velocity of the fuel increases until the first shock reflects off the center of the implosion. It then decreases until the stagnation phase during which most of the nuclear reactions are produced. At the stagnation phase, the temperatures and densities of the plasmas in the hot-spot become very high (required fuel ion temperature and areal density \cite{s_atzeni_ppfc_09, lindl_pop_04} are, respectively, $T_i$$>$10 keV and $\rho$$R$$>$0.3 g\,cm$^{-2}$ for achieving ignition) and may begin the burn of the fusion material in the central area. In this process, since the implosion core contains multiple ion species due to the fuel combinations and possible fuel-shell mixing during the implosion, the pressure and temperature gradients in the imploding target can drive ion species separation \cite{Amendt10prl, Amendt11pop} via interspecies diffusion (e.g., baro-, thermo-, and electro-diffusion) between the ions comprising the fuel.\cite{Kagan12prl, Kagan12pop, Kagan14pla, Kagan_xiv_16} Such a phenomenon has been proposed to explain the deficit in nuclear yield or yield-ratio anomalies reported in several ICF experimental campaigns.\cite{Bellei13pop, Rygg06pop, Wilson08jpcs, Dodd12pop, Casey12prl, Herrmann09pop, Rosenberg15pop, Rinderknecht14pop, Rinderknecht15prl} However, all these campaigns relied on comparing total yield measurements with standard radiation-hydrodynamic simulations that do not model multi-ion-species physics. To assess the effect of interspecies ion separation in ICF implosions quantitatively, first-principles analytic theories for multi-ion-species diffusion are developed and being implemented in standard implosion codes. \cite{Gittings08csd, Vold15pop, Hoffman15pop} Similarly, ion-Fokker-Planck\cite{Larroche12pop, Inglebert14epl} and particle-in-cell kinetic simulations\cite{Bellei13pop, Bellei14pop, Kwan15dpp} are also being used to study ICF implosions, particularly in kinetic scenarios.

The highlights of this work were previously reported in an earlier publication.\cite{hsu_16_epl} In this paper, we provide the details of how we extract temporally and spatially resolved experimental evidence of interspecies ion separation from imaging x-ray spectroscopy data. These x-ray data are expected to be less sensitive to other potential cause of yield degradation in an ICF implosion. Another purpose of this work is to complement and expand upon prior ICF experimental campaigns that relied on yield or yield-ratio anomalies as evidence for species separation. Our experimental campaign design was guided by the fact that interspecies thermo-diffusion would be strongest for species with large mass and charge difference as discussed in Kagan $et$ $al$.\cite{Kagan12prl, Kagan12pop, Kagan14pla} Therefore, we chose implosions with D$_2$/Ar fill to maximize interspecies diffusion via thermo-diffusion in order to obtain proof of interspecies ion separation. We chose Ar in particular because of its proven usefulness as an x-ray spectroscopic tracer in ICF implosions. We analyze x-ray image data recorded by Multi-Monochromatic x-ray Imager (MMI) instruments.\cite{Koch05rsi, Nagayama11jap} The MMI data are spectrally, temporally and spatially resolved and can be processed to obtain both narrow-band images and sets of space-resolved spectra, thus far obtainable only from MMI.\cite{Nagayama11jap} Analyses of the space-resolved spectra/narrow-band images yield spatial profiles of plasma electron temperature, electron density,  argon atom number density, deuterium atom number density, and argon atom number fraction in the implosion core. 

The structure of this paper is as follows. In Sec.\:II, we provide details about our OMEGA direct-drive ICF experiments. In Sec.\:III, Ar modeling and the codes to generate theoretical atomic databases are discussed. Section\:IV covers all the details of the analysis method and error analysis. Similarly, in Sec.\:V, we show spatial profiles of Ar concentration obtained from xRAGE \cite{Vold15pop} simulations with two-ion-species transport model, and use it to interpret experimentally inferred Ar concentration profiles. Finally, we summarize the paper with some ideas to improve the current analysis method in the future.
\begin{figure}[h!]
\centering
\includegraphics[width=5cm]{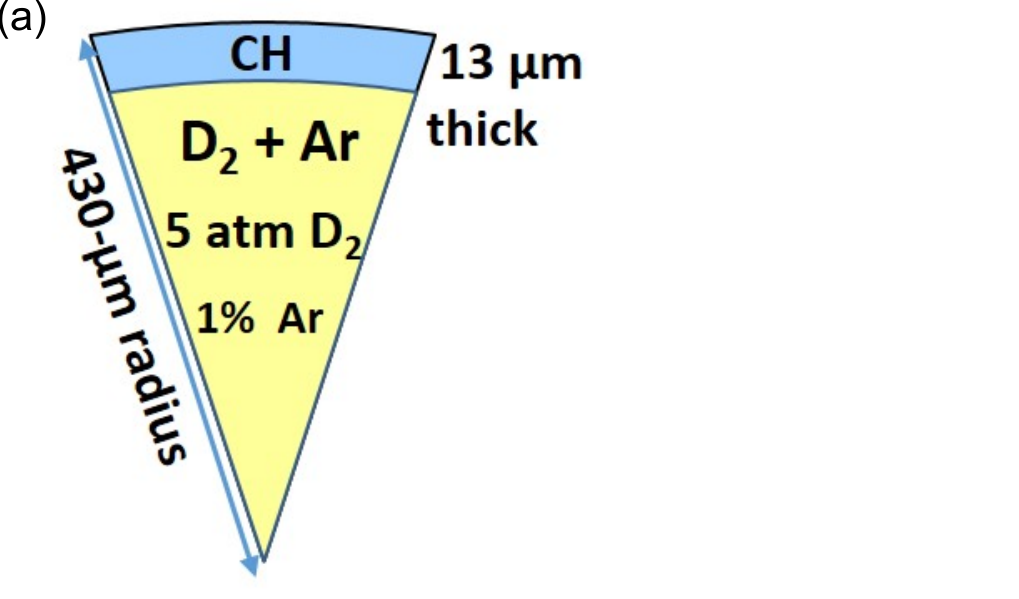}
\includegraphics[width=5cm]{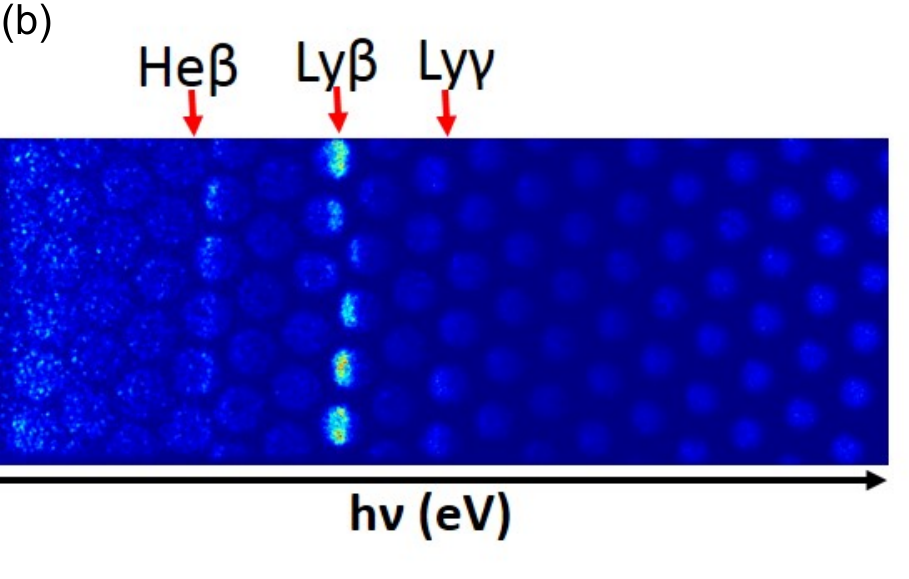}
\caption{(a) A diagram of the spherical target used in OMEGA shot 78199 (shell thickness: 13 $\mu$m; D$_2$ pressure: 5 atm; initial Ar concentration: 1$\%$).
(b) 78199 TIM 3 Frame 2 MMI data after processing.}
\end{figure}

\section{Experiments at OMEGA}
The direct-drive ICF experiments were performed on the OMEGA laser facility\cite{Boehly97oc} of the Laboratory for Laser Energetics at the University of Rochester, NY using 60 laser beams. All laser beams were configured to have the same energy, and delivered a total amount of about 21.5 or 26 kJ of UV laser energy on target with a 1 ns square pulse shape. The targets were spherical plastic shells (outer radius $\approx$ 430--439 $\mu$m) of either 13-$\mu$m shell thickness with 5-atm D$_2$/Ar gas fill or 15-$\mu$m shell thickness with 3-atm D$_2$/Ar gas fill. For each thickness/fill combination, we had both 1.0$\%$ and 0.1$\%$ Ar by atom fraction. We use data from shot 78199 as an illustrative case in this paper. For this shot, the total laser energy on target, outer diameter, shell thickness, D$_2$/Ar gas fill and Ar fraction ($f_{Ar}$) were approximately 21 kJ, 430 $\mu$m, 13 $\mu$m, 5 atm and 1$\%$ (by atom), respectively.

The Ar x-ray signal from the core was primarily observed during the deceleration phase of the implosion and recorded with a streaked crystal x-ray spectrometer (mounted on TIM 1, where TIM stands for ``ten-inch manipulator" diagnostic port), a time- and space-integrated, absolutely calibrated x-ray spectrometer (mounted on TIM 2) and two gated, MMI instruments with quasi-orthogonal views mounted on TIM 3 and TIM 4. Standard neutron diagnostics (12-m neutron time-of-flight scintillator and crogenic neutron temporal diagnostic) and full-aperture backscatter systems were also fielded. For our case of illustration, the yield (DD-n) and burn-weighted ion temperature were 1.35$\times$$10^{11}$ and 6.48 keV, respectively, obtained from the 12-m neutron time-of-flight scintillator. Similarly, the neutron bang time and burn width were 1.313 ns and 116 ps, respectively, obtained from  the cryogenic neutron temporal diagnostic (cryo-NTD). The streaked crystal spectrometer recorded temporally and spectrally resolved but spatially integrated x-ray data from the core. The MMI instrument consists of a pinhole array, a multi-layered mirror (MLM), and a micro-channel-plate (MCP) framing camera detector. In particular, the framing cameras were equipped with 4 MCPs thus providing the opportunity for recording up to four snapshots of the implosion with a temporal resolution of approximately 100 ps. The pinhole array consists of around 1000 pinholes approximately 10 $\mu$m in diameter with an average separation of 115 $\mu$m, setting the spatial resolution of the instrument at $\approx$ 10 $\mu$m.\cite{Nagayama11jap} The MLM (consists of 300 bilayers of tungsten and boron-carbide with an average thickness of 15 \AA\ of each) reflects the photons coming from the pinholes towards the MCP according to Bragg's law.\cite{Nagayama11jap} The spectral resolution power of the MMI is $\lambda/\Delta\lambda$$\approx$150.\cite{Nagayama11jap} The MMI recorded spectrally resolved implosion core images on four MCPs i.e. four frames (frame 1 is the earliest and frame 4 is the latest in time) between 3.3-5.5 keV around the time of the first shock convergence. Figure\:1a shows the pie-slice diagram of the spherical target used in the OMEGA shot 78199, and Fig.\:1b shows 78199 TIM 3 Frame 2 MMI data after processing. The horizontal axis represents the photon energy increasing from left to right measured in units of eV, and vertical axis represents an array of implosion core pinhole images. The MMI data shown in Fig.\:1b consist of Ar K-shell line transitions, i.e., Ar He$\beta$ ($1s^2-1s3p$, 3680 eV), Ly$\beta$ ($1s-3p$, 3940 eV), Ly$\gamma$ ($1s-4p$, 4150 eV), and x-ray continuum radiation. Note that the He$\gamma$ ($1s^2-1s4p$, 3875 eV) is blended into the red wing of Ly$\beta$. We obtained analyzable MMI data for several shots with initial $f_{Ar} =1 \%$. Shots with initial $f_{Ar} = 0.1\%$ provided weaker spectral lines as predicted\cite{hsu_16_epl} and not suitable for analysis. The important steps of processing of raw MMI data include conversion of film density to intensity using film calibration, artifact removal and flat-fielding of the data, center determination of monochromatic core images, photon-energy dependent intensity corrections associated with beryllium-filter transmission, reflectivity of the multilayered Bragg mirror, and spectral response of the MCP.\cite{Nagayama11jap, Rochau_rsi6} The details about the MMI instrument, data processing method, and the extractions of narrowband images and space-resolved spectra are described elsewhere.\cite{Nagayama11jap, Nagayama14pop, Nagayama15rsi} 
\begin{figure}[h!]
\centering
\includegraphics[width=8cm]{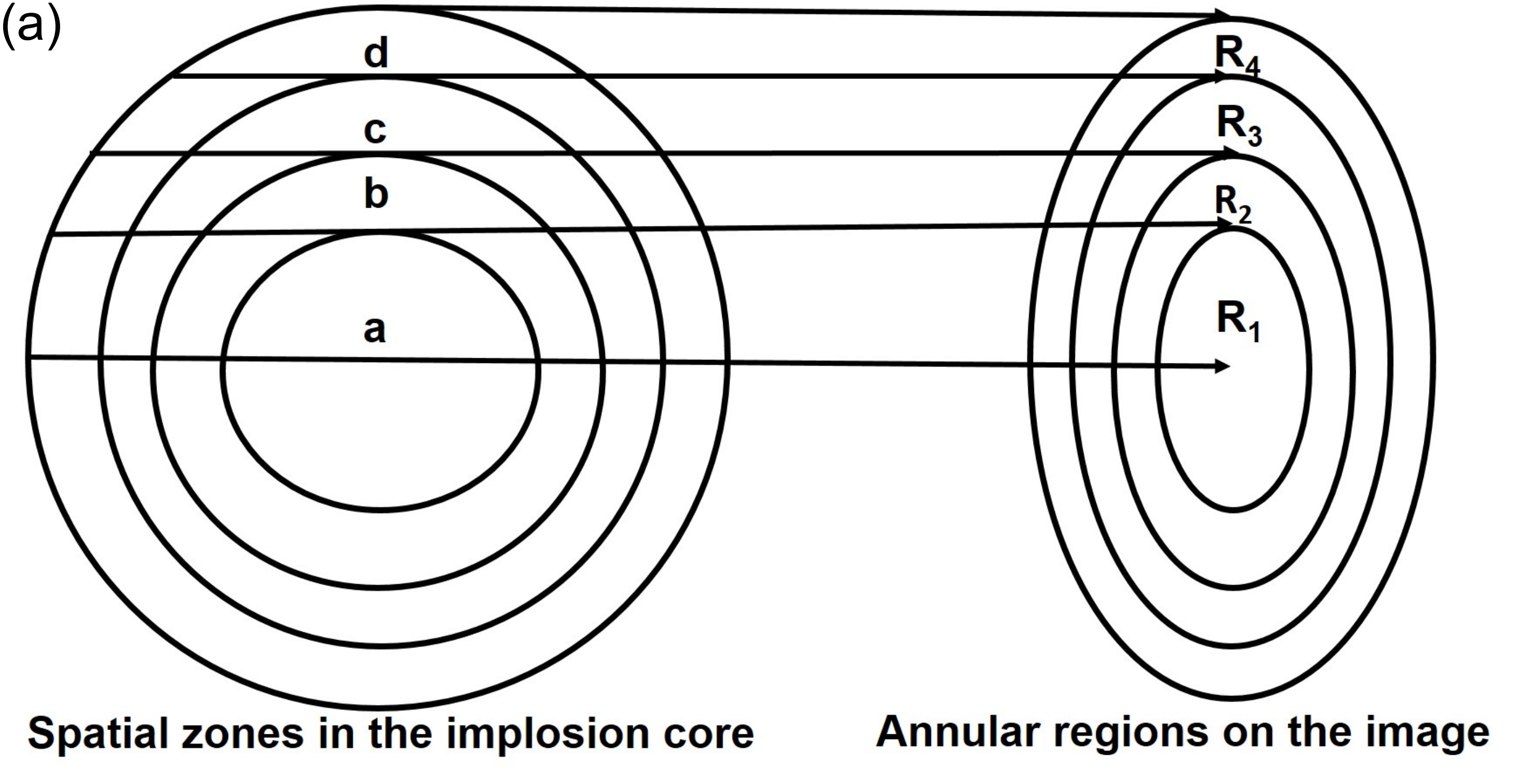}\\
\vspace{0.05cm}
\includegraphics[width=8cm]{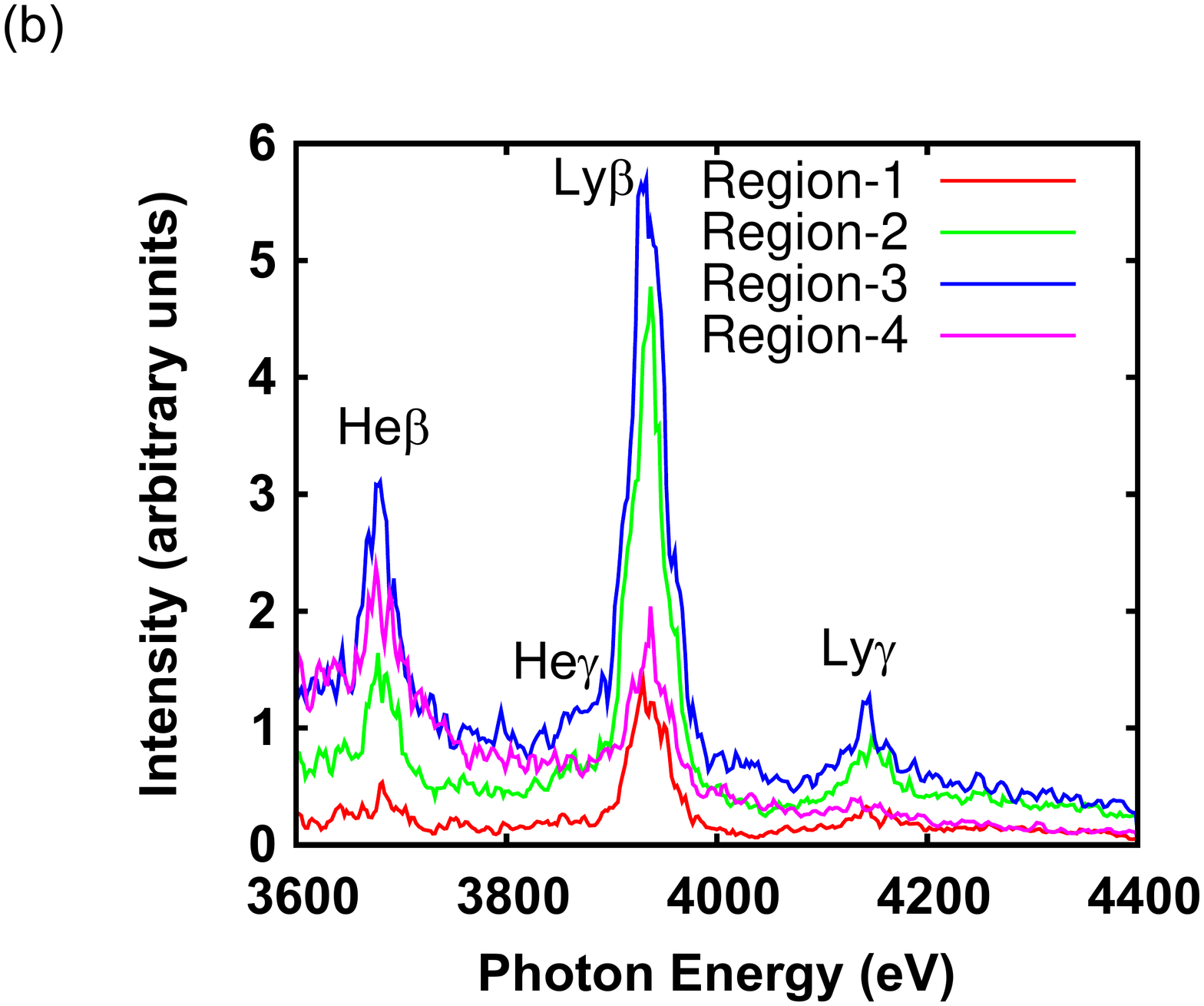}
\caption{(a) Schematic illustration of spherical shell zones in the implosion core and annular regions on the core image. (b) Space-resolved spectra extracted from the annular regions.}
\end{figure}
For this work, we reconstructed narrowband images of Ar He$\beta$, Ly$\beta$ and Ly$\gamma$ from the MMI data. Similarly, space-resolved spectra were extracted from the annular regions on the core images across the MMI data. The selection of the annular regions must be consistent with the following two requirements: (1) size larger than spatial resolution of the MMI instrument ($\approx$10 $\mu$m), (2) good signal-to-noise ratio. For the data discussed in this paper, we have found that the images can be divided into four annular regions that are consistent with these requirements. These annular regions on the core image can be considered as the projection of four concentric spherical shell zones in the implosion core. The sizes of these zones are given by the details of the annular regions selected on the image. Figure\:2a shows the schematic of the spatial zones (represented by ``a", ``b", ``c" and ``d") in the implosion core and their projections on the image plane (annular regions are represented by ``$R_1$", ``$R_2$", ``$R_3$" and ``$R_4$") and Fig.\:2b shows the space-resolved spectra extracted from the annular regions on the spectrally-resolved core images of the OMEGA shot 78199 TIM 3 Frame 2 data shown in the Fig.\:1b.

\section{Argon Modeling}

\begin{figure*}
\includegraphics[width=12cm]{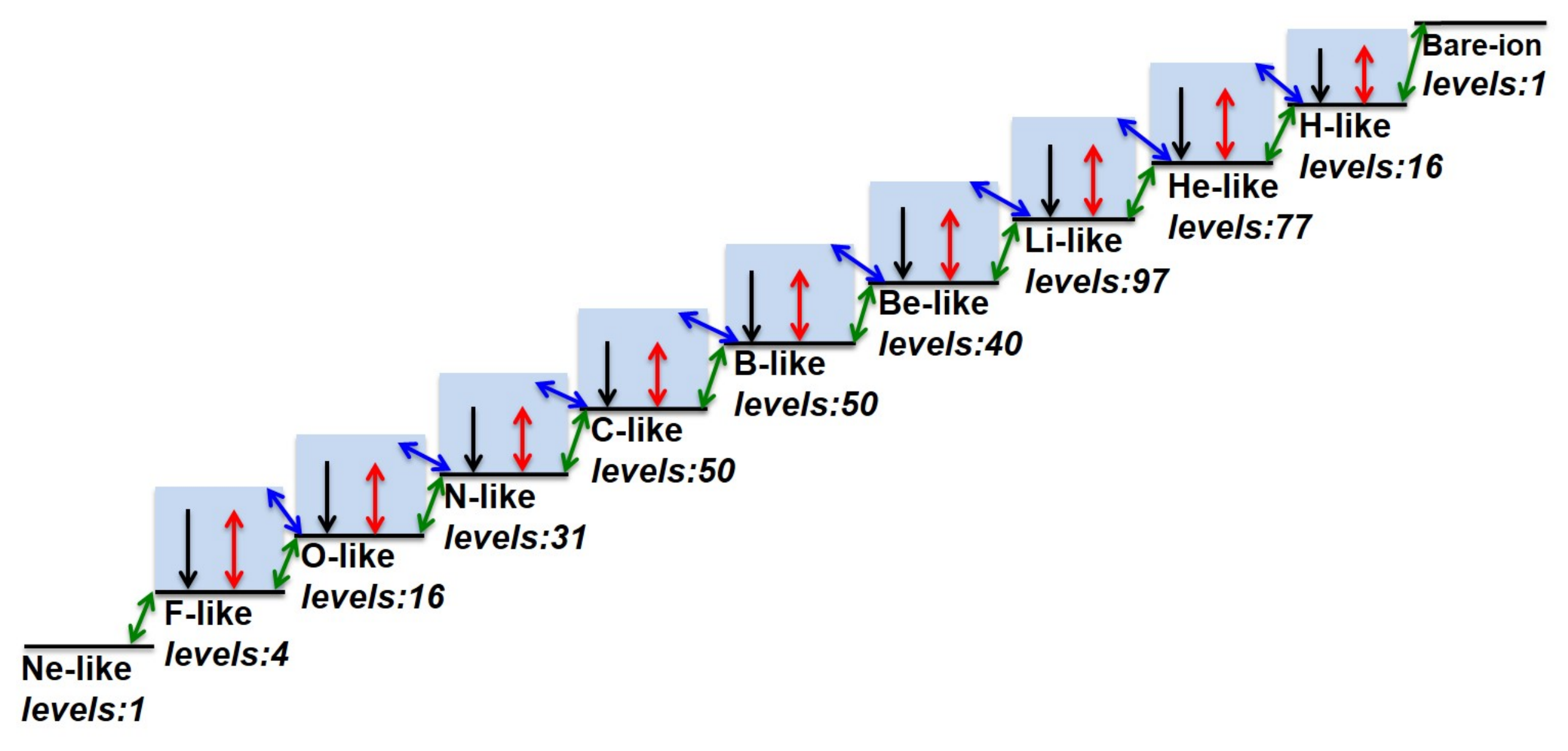}
\caption{Argon atomic structure and ATOMIC kinetics model. Black arrow is for radiative decay, red for electron excitation and de-excitation, green for photoionization, electron ionization and their inverse processes, and blue for autoionization and process of electron capture.}
\end{figure*}
The calculation of x-ray spectral emissions was carried out with the Los Alamos suite of codes.\cite{Fontes15jpb} Atomic wave functions, oscillator strengths, and plane-wave Born collisional excitation cross sections were obtained from the fully relativistic code RATS, which is based on the work of Sampson $et$ $al$.\cite{Sampson09pr} The selection of ion stages and electronic configurations was informed by the temperature and density conditions achieved in our experiments.\cite{Hakel14pop} Our Ar model considers all ionization states from Ne-like Ar to the fully-stripped ion and includes 383 energy levels (fine-structure), with the following maximum number of energy levels per ion: 1 fully stripped, 16 H-like, 77 He-like, 97 Li-like, 40 Be-like, 50 B-like, 50 C-like, 31 N-like, 16 O-like, 4 F-like and 1 Ne-like Ar. A key subset of the plane-wave Born excitation cross sections from RATS was superseded with the more accurate distorted-wave cross sections calculated by the ACE code.\cite{Fontes15jpb, Clark88} Data for collisional ionization, autoionization, and photoionization were computed with the GIPPER code.\cite{Fontes15jpb, clark91apj} Monochromatic emissivity, absorption, and scattering databases for the elements present in our experiments were calculated from the above-described data via a steady-state, collisional-radiative equilibrium model implemented in the ATOMIC code.\cite{Fontes15jpb, Fontes16, Abdallah90} Finally, the FESTR code\cite{Hakel16cpc} draws on the ATOMIC-generated databases to produce spectroscopic-quality x-ray outputs by the postprocessing of hydrodynamics simulations, and for the verification of our analysis technique (see Sec.\:IV E). Figure\:3 shows ion stages, total maximum number of fine-structure levels per ion and different atomic processes (represented by colored arrows) included in our model.

\section{Analysis Method}
Our analysis method to infer Ar atom number fraction spatial profile consists of three steps. The first step is to extract electron temperature ($T_e$) and density ($n_e$) spatial distributions in the implosion core by using an emissivity analysis method.\cite{Welser-Sherrill07pre} In the second step of the analysis, we extract spatial distributions of Ar atom number density ($n_{Ar}$) by using an inversion method based on the proportionality between intensity of a line emission and corresponding upper-level population density.\cite{Joshi15} The final step uses the charge quasi-neutrality constraint to obtain the deuterium number density ($n_D$) and Ar atom number fraction ($f_{Ar}$). We describe all the steps in detail in the following paragraphs.

\subsection{Extraction of ${T_e}$ and $n_e$}
In this step, we start from the extraction of area-normalized intensities from the annular regions in the narrow-band images. We divide each image into four annular regions that are consistent with the requirements mentioned in the experimental section. Figure\:4a shows the continuum-subtracted He$\beta$, Ly$\beta$ and Ly$\gamma$ narrow-band images reconstructed from the MMI data of the OMEGA shot 78199 TIM 3 Frame 2. The spectral ranges used for the reconstruction of the narrow-band images are 3640-3720 eV for He$\beta$, 3895-3975 eV for Ly$\beta$, and 4110-4190 eV for the Ly$\gamma$; each has a bandwidth of 80 eV. Figures\:4b and 4c show area-normalized intensity spatial profiles extracted from the annular regions on the images and corresponding Abel-inverted emissivity profiles in the implosion core. We use a generalized Abel inversion (assumes optically thin approximation),\cite{Welser-Sherrill07pre} in which the intensity of each pixel on the image plane is associated with the line integral of emissivity along a chord in the core. The lines Ly$\beta$, He$\beta$ and Ly$\gamma$ are normally considered as optically thin.\cite{Welser-Sherrill07pre} The three necessary conditions to apply Abel inversion in our intensity profiles are: 1) the intensity must be zero at the outermost region and 2) the first derivative of the intensity with respect to the radius should vanish at the central region, and 3) the narrow-band images must be sufficiently symmetric. To achieve the first criterion in our intensity profiles, we make extrapolations of the intensities at the outermost regions on the images. For the second criterion, if our profile is not fairly flat near the central region, we replace the central value by the value next to it. We will discuss the symmetry of narrow-band images later in the paper (Sec.\:IV D).
\begin{figure}[h!]
\centering
\includegraphics[width=8cm]{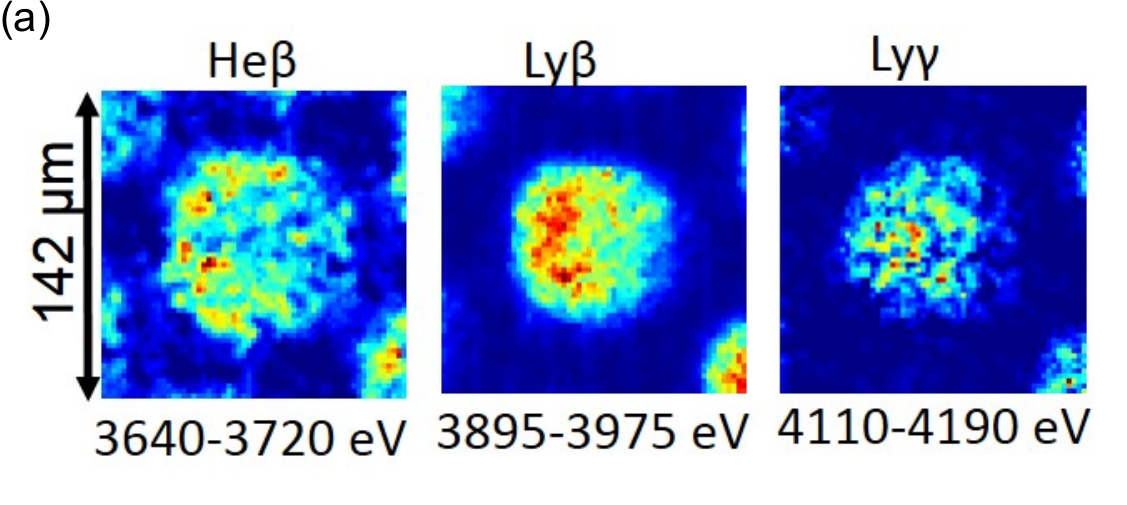}\\
\vspace{-0.25cm}
\includegraphics[width=8.5cm,height=6.5cm]{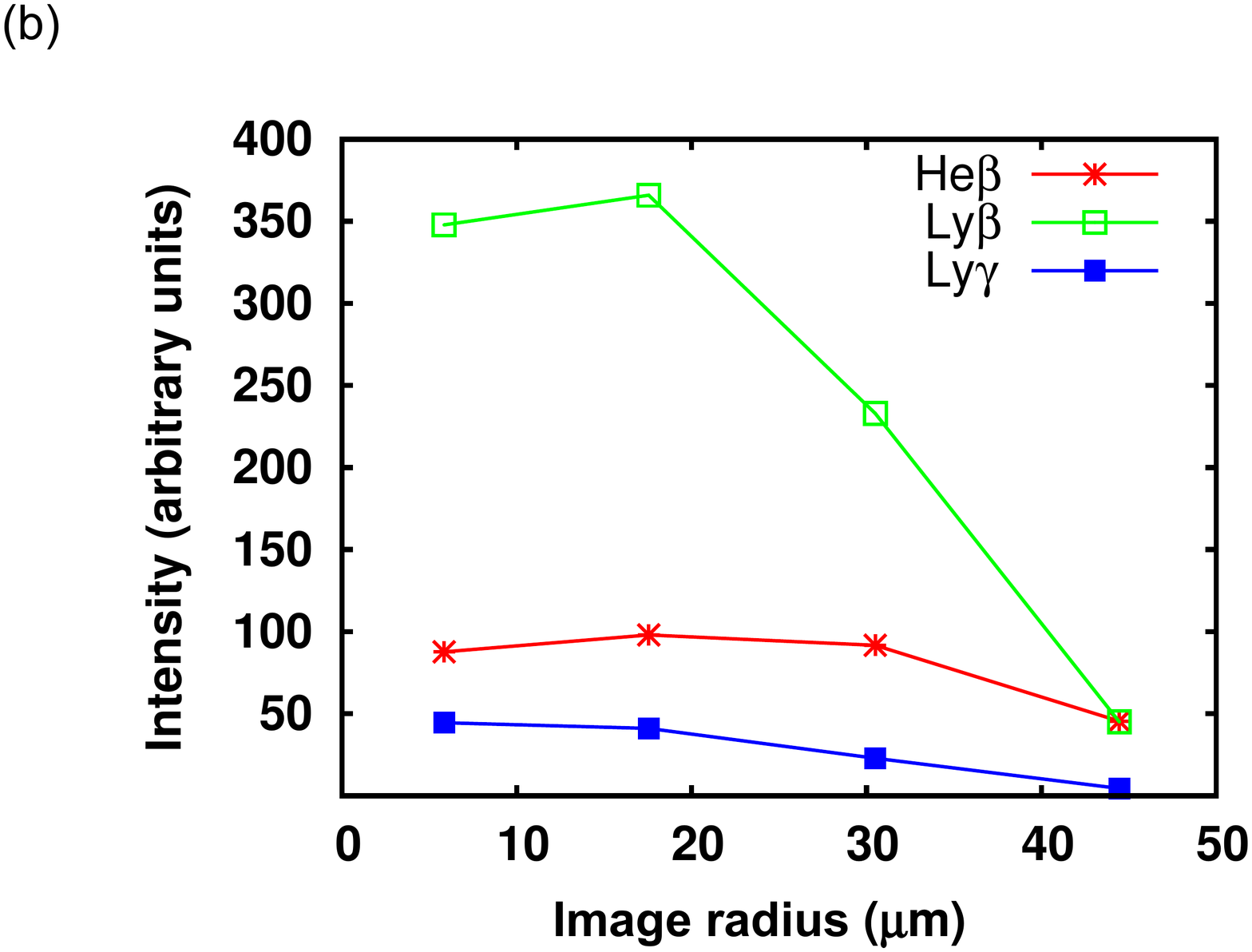}\\
\vspace{-1cm}
\includegraphics[width=8cm,height=6.5cm]{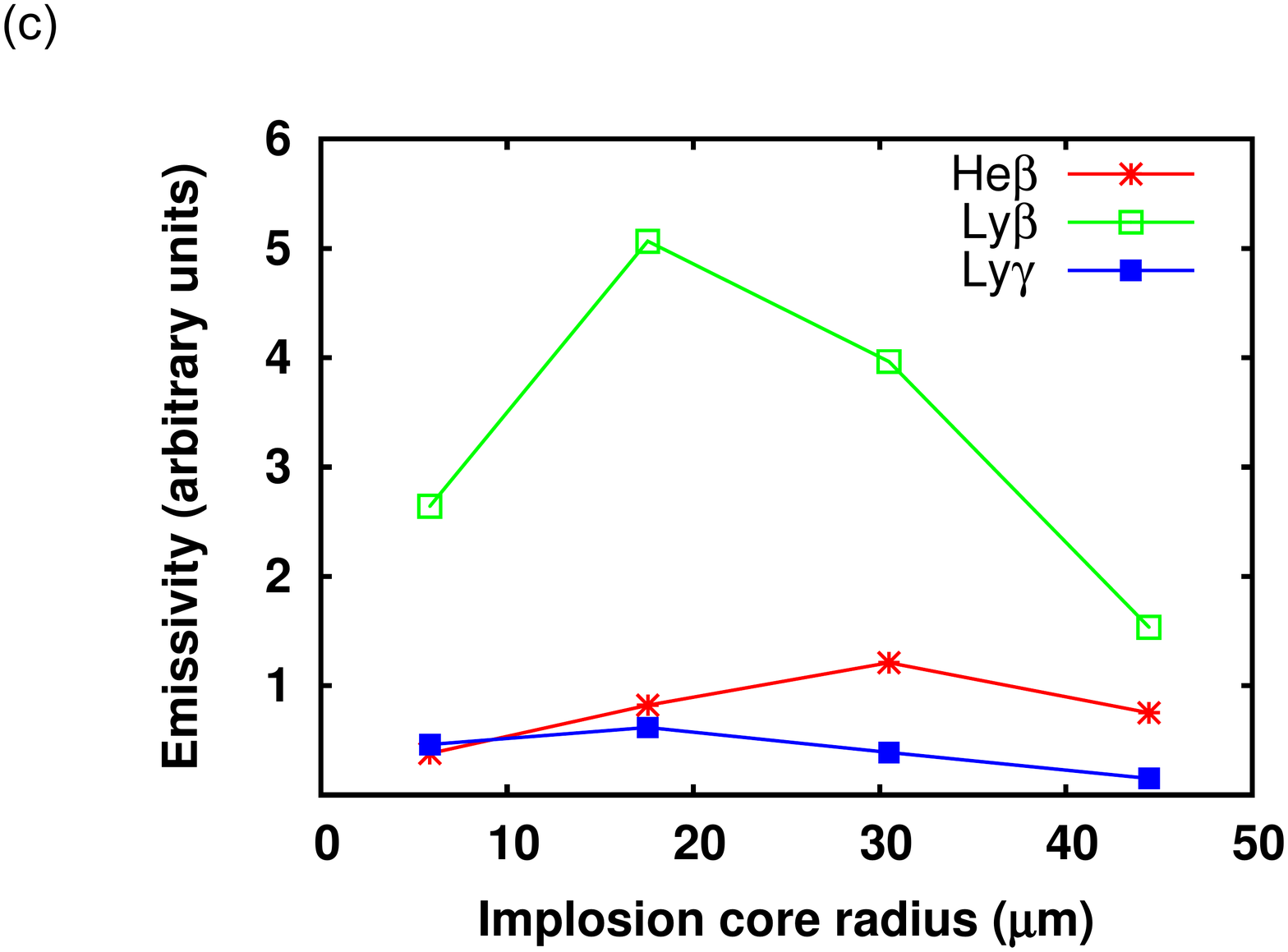}
\caption{(a) Ar He$\beta$, Ly$\beta$ and Ly$\gamma$ narrow-band images after continuum subtraction. Energy ranges shown under the images are used to reconstruct the images from MMI data. (b) Area-normalized radial line-outs, and (c) corresponding emissivity profiles for the He$\beta$, Ly$\beta$ and Ly$\gamma$ narrow-band images.}
\end{figure}

The emissivity of a line emission strongly depends on the electron temperature and density. But the emissivity ratio of Ly$\beta$/He$\beta$ is strongly dependent on the electron temperature and weakly dependent on the electron density.\cite{Welser-Sherrill07pre} We compute an array of theoretical emissivity ratios of Ly$\beta$/He$\beta$ as a function of electron temperature at constant electron density, and use it to relate the spatial distribution of the experimental emissivity ratio of Ly$\beta$/He$\beta$ with the spatial distribution of the electron temperature in the core.\cite{Welser-Sherrill07pre} Theoretical emissivities are generated via the Los Alamos suite of atomic codes explained in Sec.\:III. The total emissivity of a line (defined by bandwidth) used in the analysis is a superposition of many single line contributions, including resonance and satellite transitions falling in the narrow-band spectral range. To study the sensitivity of the ratio of the theoretical emissivity (Ly$\beta$/He$\beta$) on electron density, we generate an array of theoretical emissivity ratios (Ly$\beta$/He$\beta$) as a function of electron temperature at four different electron densities (5.0$\times10^{23}$, 8.0$\times10^{23}$, 1.0$\times10^{24}$, 2.0$\times10^{24} $cm$^{-3}$) in our regime of interest, and use them to extract four temperature profiles in the core. The final $T_e$ profile shown in Fig.\:5a is the average of the four temperature profiles.

Next, we extract the spatial distribution of the electron density by using the information about the electron temperature profile as obtained above, and solving the following equation, which relates the theoretical and experimental emissivity,\cite{Welser-Sherrill07pre}
\begin{equation}
\varepsilon_{Line}^{Expm}= k\varepsilon_{Line}^{Theory}(T_{e}, n_{e}),
\end{equation}
where $\varepsilon_{Line}^{Expm}$, $\varepsilon_{Line}^{Theory}$ are the experimental and theoretical emissivities of a narrow-band line, and $k$ is the scale factor between the experimental (in arbitrary units) and theoretical emissivities (in absolute units). There are two unknowns in the equation, i.e., $n_e$ and the scale factor $k$. We fit the experimental space-integrated spectrum (spectral range 3600-4200 eV; which includes Ly$\beta$, He$\beta$ and Ly$\gamma$) to find the spatially averaged density condition in the core by assuming the implosion core as a single uniform slab in the optically thin approximation. We then use the spatially averaged density to determine the scale factor between the experimental and the theoretical emissivity in Eq.\:(1), and solve the equation to extract the $n_e$ gradient in the implosion core. Next, an emissivity-weighted average of the electron density gradient is computed.\cite{Welser-Sherrill07pre} If it is equal or very close to the spatially averaged electron density extracted from the fitting of space-integrated spectrum, the density gradient is considered to be the solution to the above equation. If not, the scale factor $k$ (and the $n_e$ gradient obtained by solving Eq.\:(1)) is again computed with slightly higher and lower electron densities than the spatially averaged density mentioned above until the gradient profile with the closest match between the emissivity-weighted average and the spatially averaged electron density is obtained. Figure\:5b shows spatial profiles of $n_e$ obtained from all three lines and their average.
\begin{figure}[h!]
\centering
\includegraphics[width=8cm,height=6.5cm]{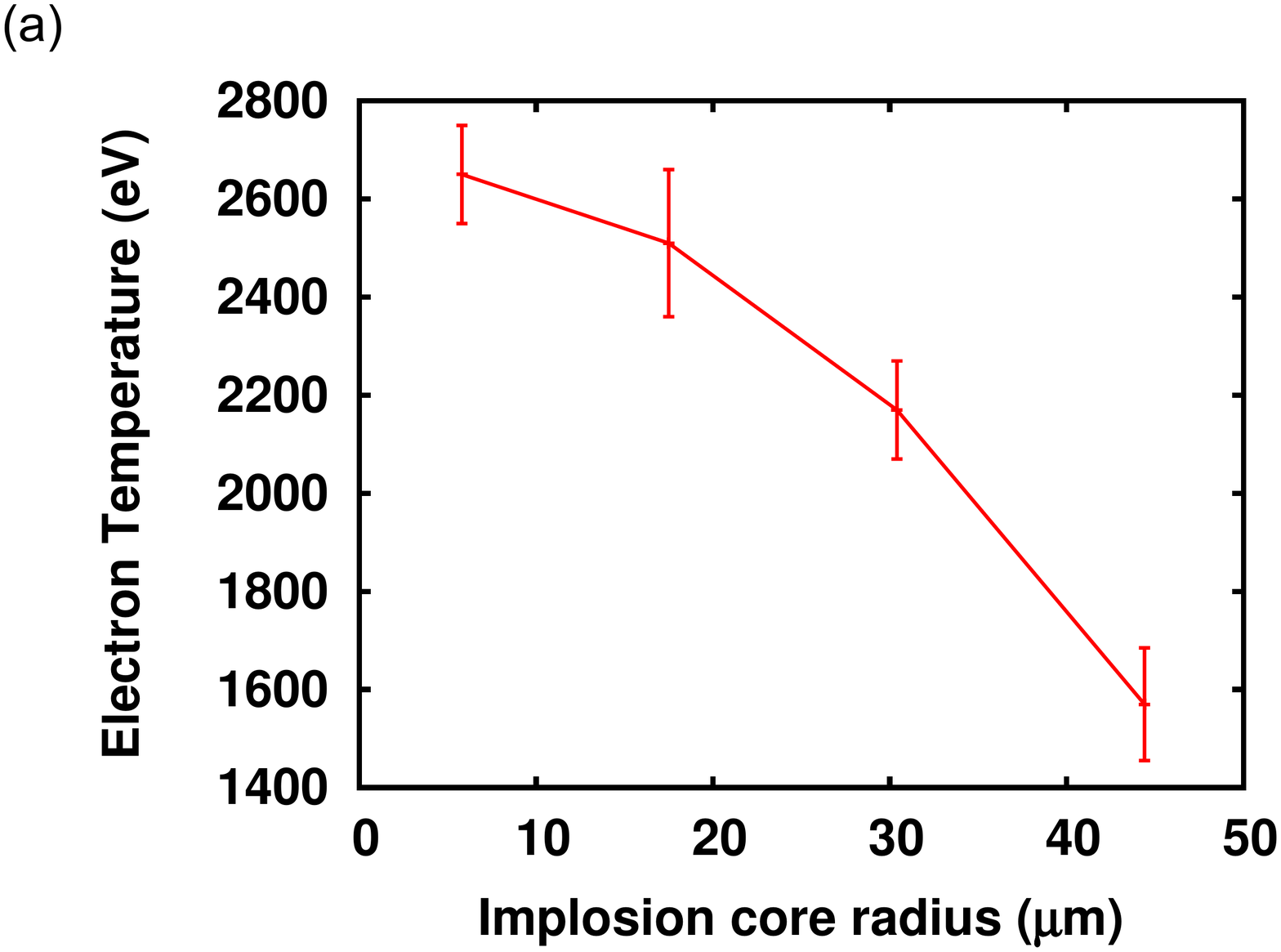}\\
\vspace{-0.75cm}
\includegraphics[width=8.9cm,height=6.5cm]{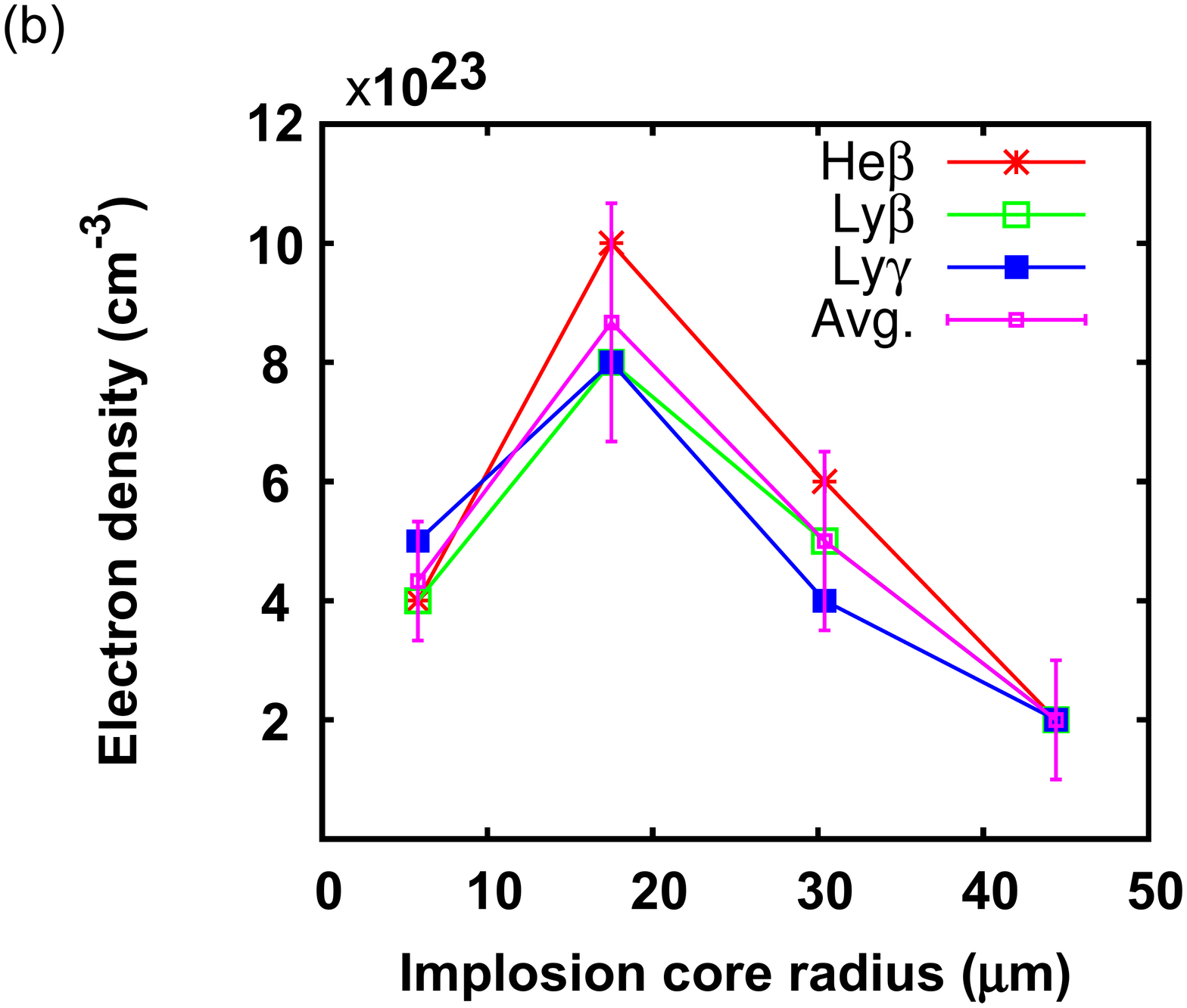}
\caption{(a) Electron temperature and (b) density spatial profiles from the analysis of shot 78199 TIM 3 Frame 2}
\end{figure}

\subsection{Extraction of $n_{Ar}$}
In the second step of the analysis, we start with the extraction of spatial distributions of upper-level populations associated with different K-shell line emissions of Ar ions (in arbitrary units) in the implosion core, and use them to extract ion number densities (in absolute units).\cite{Joshi15} To extract spatial profiles of upper-level population densities, we exploit the proportionality of the total line intensity (i.e., lineshape integrated) in the optically thin approximation to the upper-level population number density, the photon energy, the spontaneous radiative decay of the line transition and the volume of the emission. A system of equations is set up for the total intensities of a line in all four annular space-resolved spectra. The solution of the equations provides the set of the upper-level population number densities (in arbitrary units) associated with a line emission in all four spatial zones in the core (spatial zones are shown in the Fig.\:2a),                                                
\begin{subequations}
\begin{align}
&I_{4} = \frac{1}{4\pi }n_{ud}A_{ul}h\nu_{ul} V_{d}\\
&I_{3} = \frac{1}{4\pi }(n_{uc}V_c+n_{ud}V_{d^{\prime}})A_{ul}h\nu_{ul}\\
&I_{2} = \frac{1}{4\pi }(V_bn_{ub}+V_{c^{\prime}}n_{uc}+V_{d{\prime\prime}}n_{ud})A_{ul}h\nu_{ul}\\
&I_{1} = \frac{1}{4\pi }(V_an_{ua}+V_{b^{\prime}}n_{ub}+V_{c^{\prime\prime}}n_{uc}+V_{d^{\prime\prime\prime}}n_{ud})A_{ul}h\nu_{ul} 
\end{align}
\end{subequations}
where $I_4$, $I_3$, $I_2$ and $I_1$ are the total line intensities obtained from the space-resolved spectra extracted from the annular regions $R_4$, $R_3$, $R_2$ and $R_1$, respectively, of the implosion core image. $A_{ul}$ and $h\nu_{ul}$  are the spontaneous radiative decay rate and photon energy associated with the center of a line, which we obtained from the LANL atomic database. $``V"$ are the volumes of different spherical sections (which define domains of integrations in the core for annular regions in the image) of the implosion core, and obtained from the MMI processing code\cite{Nagayama11jap} and geometry analysis. $n_{ua}$, $n_{ub}$, $n_{uc}$, and $n_{ud}$ are the upper-level population number densities of a line in the spherical zones a, b, c and d, respectively, of the implosion core (as shown in the schematic of the implosion core in Fig.\:2a). Total line intensities are obtained from the experimental space-resolved spectra extracted from the annular regions of the implosion core image by numerically integrating the area under the line after continuum subtraction. Figure\:6a shows an example to illustrate the method to obtain continuum-subtracted total intensity associated with line emissions. Integrated regions are shown as shaded. For each line, total intensities are found from all four space-resolved spectra. The narrow-band spectral range is always the same for a chosen spectral line. Equation\:(2a) outputs the upper-level population density in the outermost spherical zone, i.e., $d$ as shown in Fig.\:2a. Equation\:(2b) uses the output of Eq.\:(2a) as the input and yields the upper-level population density in the spherical zone $c$. Similarly, Eq.\:(2c) uses the outputs from Eqs.\:(2a) and (2b) as the inputs, and yields the upper-level population density in the spherical zone $b$. Finally, the upper-level population densities produced by Eqs.\:(2a), (2b) and (2c) are used as the inputs to Eq.\:(2d), and which, in turn, provides the upper-level population density in the spherical zone $a$, i.e., the innermost spherical zone of the implosion core. The spectral ranges used for the analysis were 3640--3720 eV, 3895--3975 eV and 4110--4190 eV for He$\beta$, Ly$\beta$ and Ly$\gamma$, respectively, as in the narrow-band images. To account for the opacity effect on these intensities, the optically thick formula for the emergent intensity of a slab can be integrated with respect to photon energy to obtain an optically thick approximation to the total line intensity (Appendix B of Ref. 45). The optically thick line intensities are smaller than that obtained in the optically thin approximation due to the self-absorption of the line intensities. As a first step towards the opacity corrections, we have approximated the opacity correction in our intensities (given by Eqs.\:(2a)--(2d)) by considering an effective volume of integration (effective volumes are slightly smaller than the actual volumes)\cite{Joshi15}. In this manner we can approximate the effect of opacity while using a procedure derived within the optically thin approximation (Eqs.\:2). 

\begin{figure}[h!]
\centering
\includegraphics[width=9.2cm,height=6.5cm]{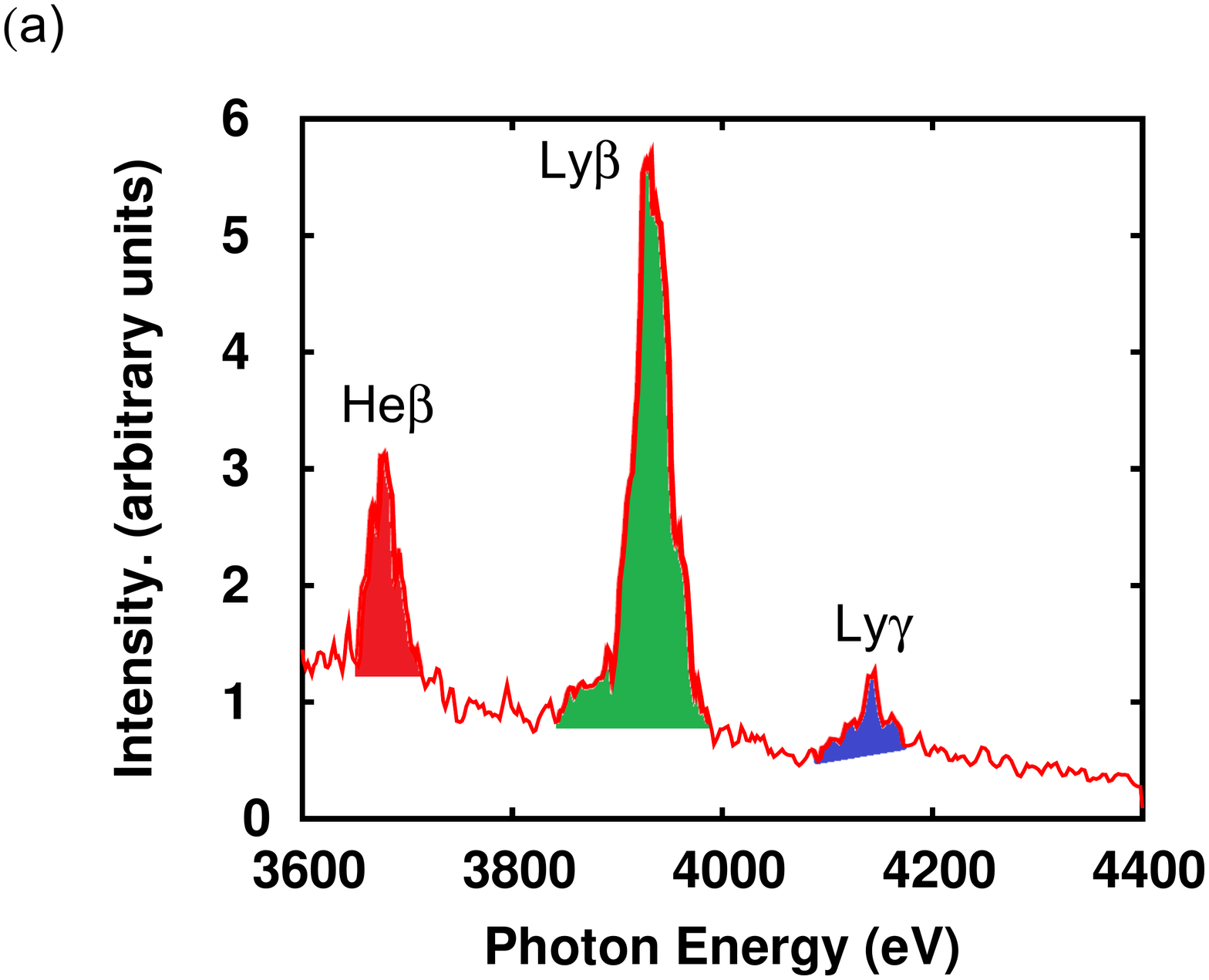}\\
\vspace{-0.5cm}
\includegraphics[width=8cm,height=6.5cm]{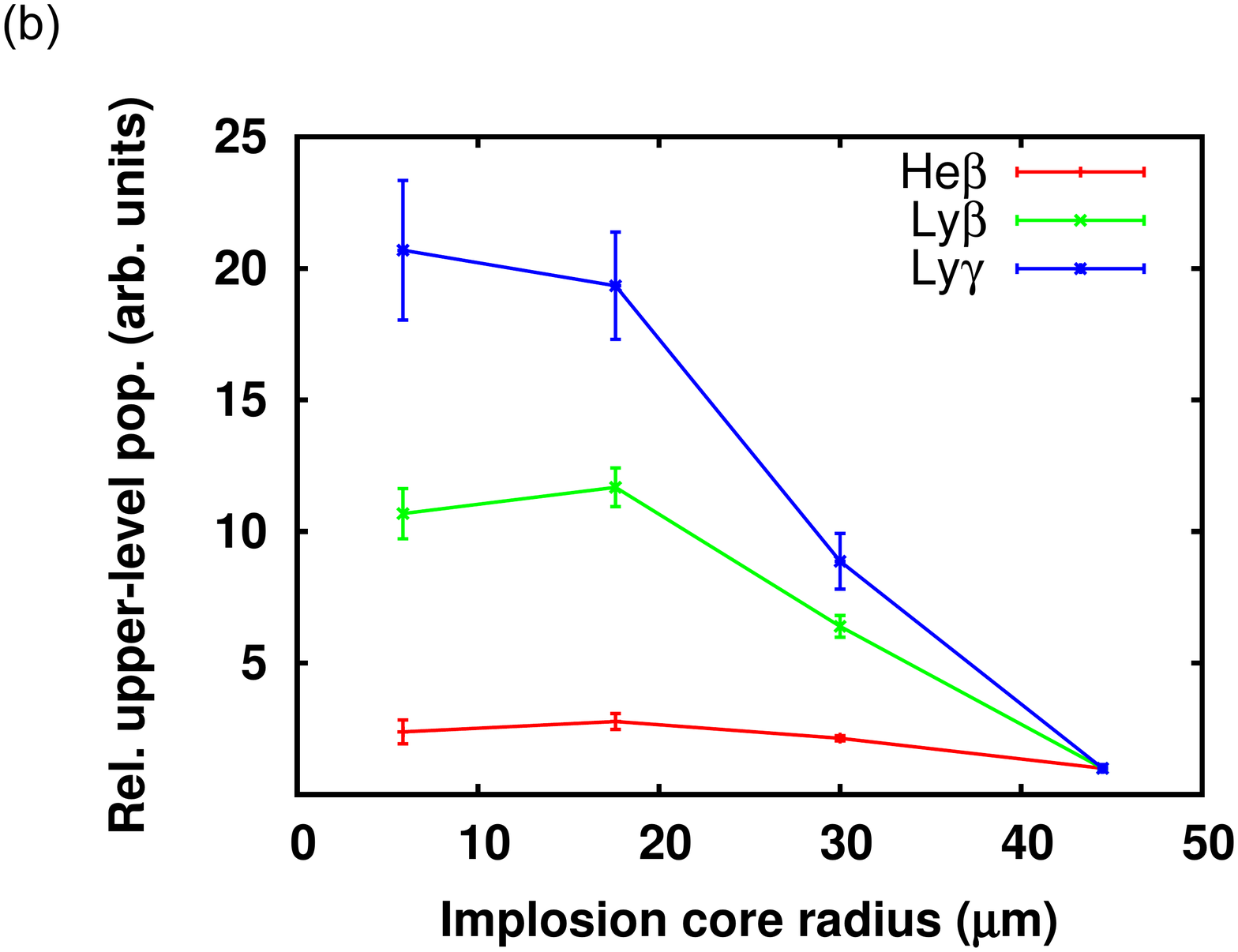}\\
\vspace{-0.5cm}
\includegraphics[width=8.8cm,height=6.5cm]{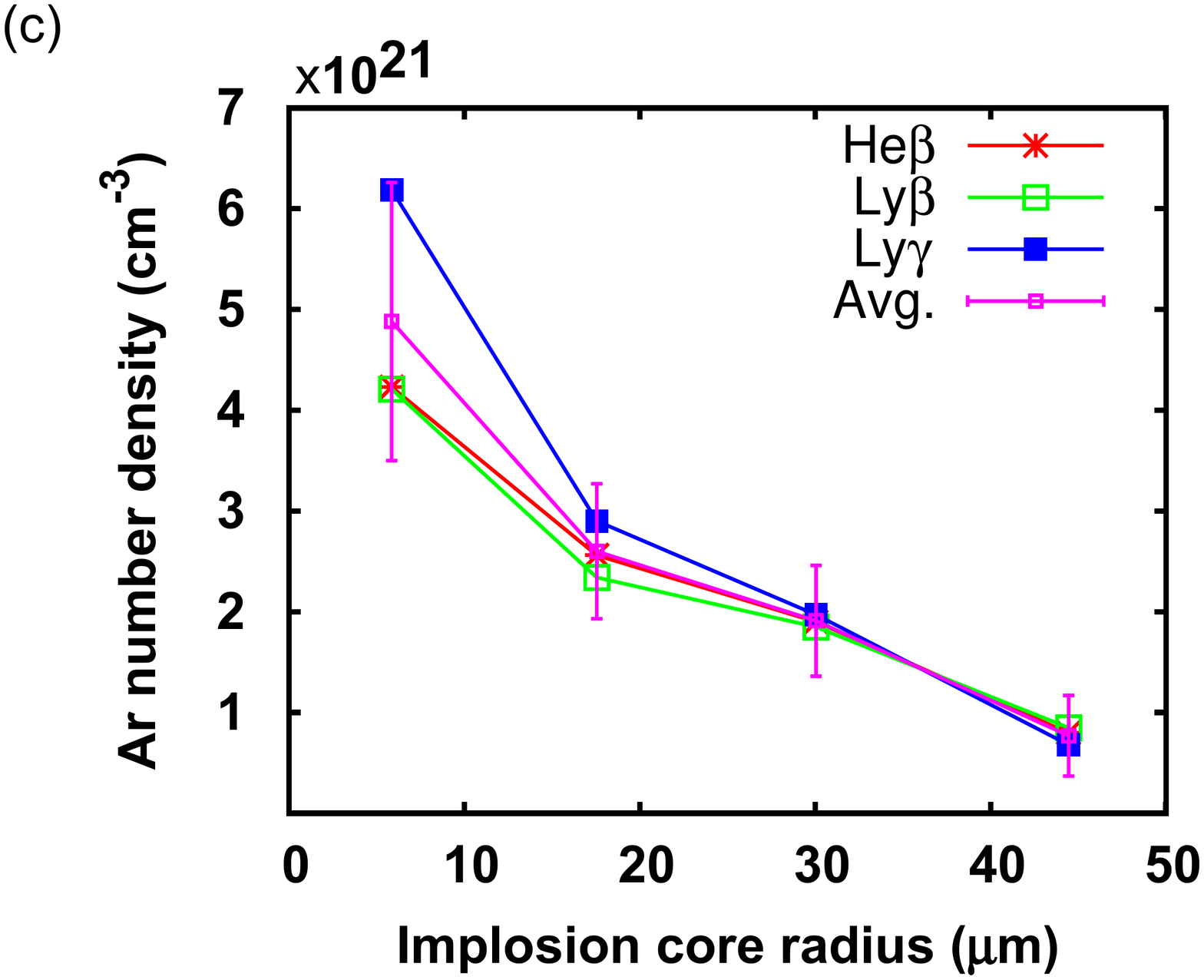}
\caption{(a) Continuum-subtracted total line intensities. For each line, total intensities are found from all four space-resolved spectra. Integrated regions are shown as shaded.  (b) Upper-level population density (normalized with respect to the values in the outermost region) vs. implosion core radius (c) Ar atom number density vs. implosion core radius extracted from the He$\beta$, Ly$\beta$ and Ly$\gamma$ in the implosion core of the OMEGA shot 78199 TIM 3 Frame 2.}
\end{figure}
To calculate the Ar ion number density, we use the following relation between the upper-level population number density of a line transition, ion number density and fractional level population of the upper-level associated with a line transition,
\begin{equation}
n_{u} = n_{Ar} F_{u}(T_{e}, n_{e})
\end{equation}
where $F_{u}(T_{e}, n_{e})$ is the fractional level population associated with the upper level of a line emission and which we retrieve from the LANL atomic database based on the previously determined (in step 1) $n_{e}(r)$ and $T_{e}(r)$ values. The upper-level population density in each spherical zone in the implosion core ($n_{u}$) is obtained as mentioned above. Ar atom number densities in arbitrary units are determined in this way from as many line emission features as observed in the data and compared to each for consistency. Finally, we use the known pre-fill amount of Ar (1$\%$ of Ar by atom, $\approx$7.45$\times$10$^{14}$ Ar atoms for shot 78199) to convert Ar atom number density in arbitrary units into absolute units by assuming conservation of Ar atoms in the implosion core. Fig.\:6b and 6c show, respectively, spatial profiles of  upper-level population density and ion number density extracted from He$\beta$, Ly$\beta$ and Ly$\gamma$ in the implosion core of OMEGA shot 78199 TIM 3 Frame 2. 

\subsection{Extraction of $f_{Ar}$}
The final step uses the charge quasi-neutrality constraint to extract deuterium number density,

\begin{equation}
n_e = n_{Ar} Z_{Ar}(T_e, n_e)+ n_D Z_D(T_e, n_e).
\end{equation}
In Eq.\:(4), everything is known except the deuterium number density ($n_D$). The ionization balances for Ar and D at the given $T_e$ and $n_e$ (as determined in step 1) are obtained from the LANL atomic database, and $n_{Ar}$ is obtained from the second step as described above. Finally, $f_{Ar}$ is inferred from the following expression:
\begin{equation}
f_{Ar} = \frac{n_{Ar}}{n_{Ar}+n_D}.
\end{equation}
Figures\:7a and 7b show spatial profiles of deuterium atom number density and argon atom number fraction extracted from the Ar He$\beta$, Ly$\beta$ and Ly$\gamma$ lines and their averages in the implosion core of the OMEGA shot 78199 TIM 3 Frame 2.

\begin{figure}[h!]
\centering
\includegraphics[width=8.9cm,height=6.5cm]{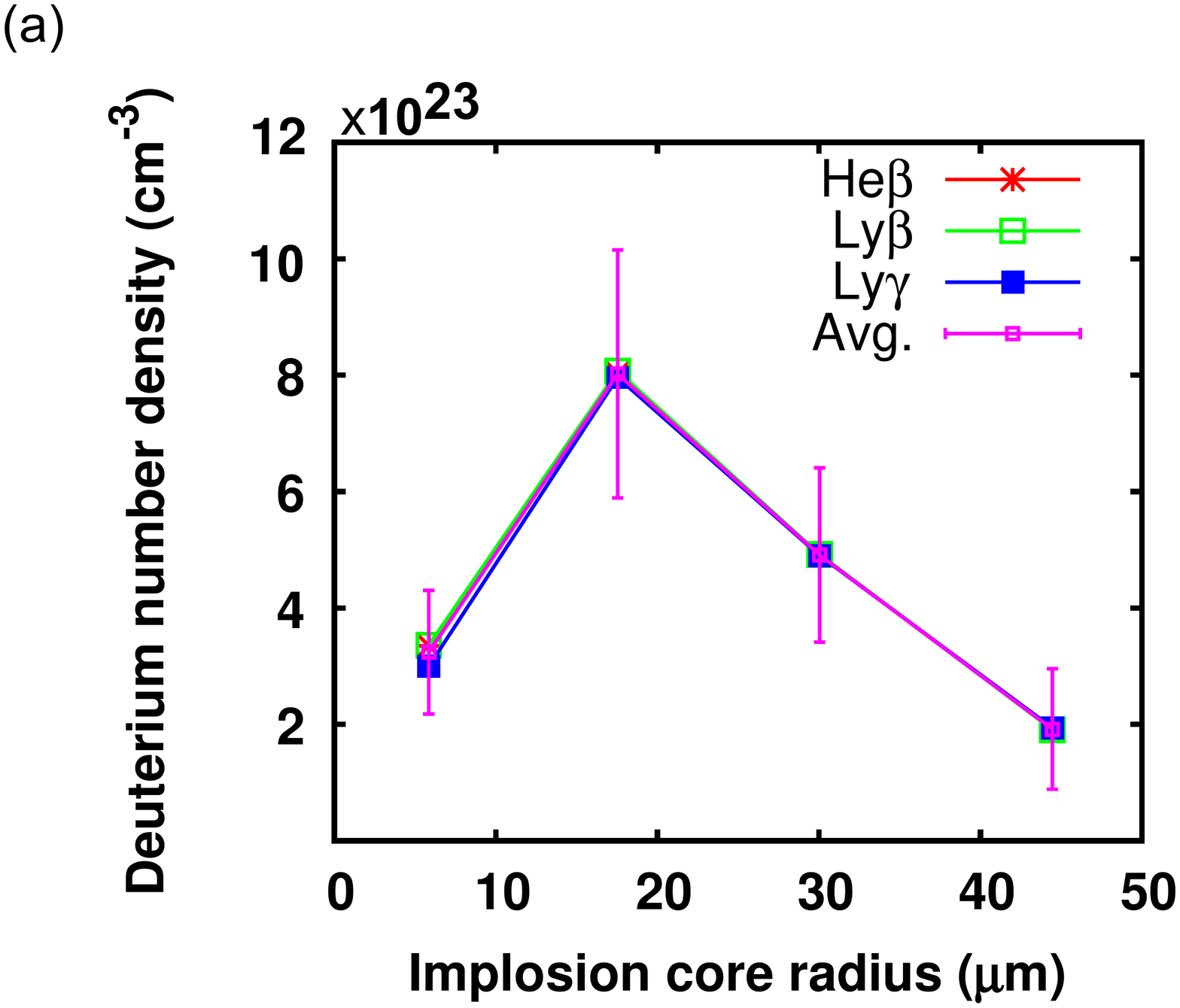}\\
\vspace{-0.5cm}
\includegraphics[width=8cm,height=6.5cm]{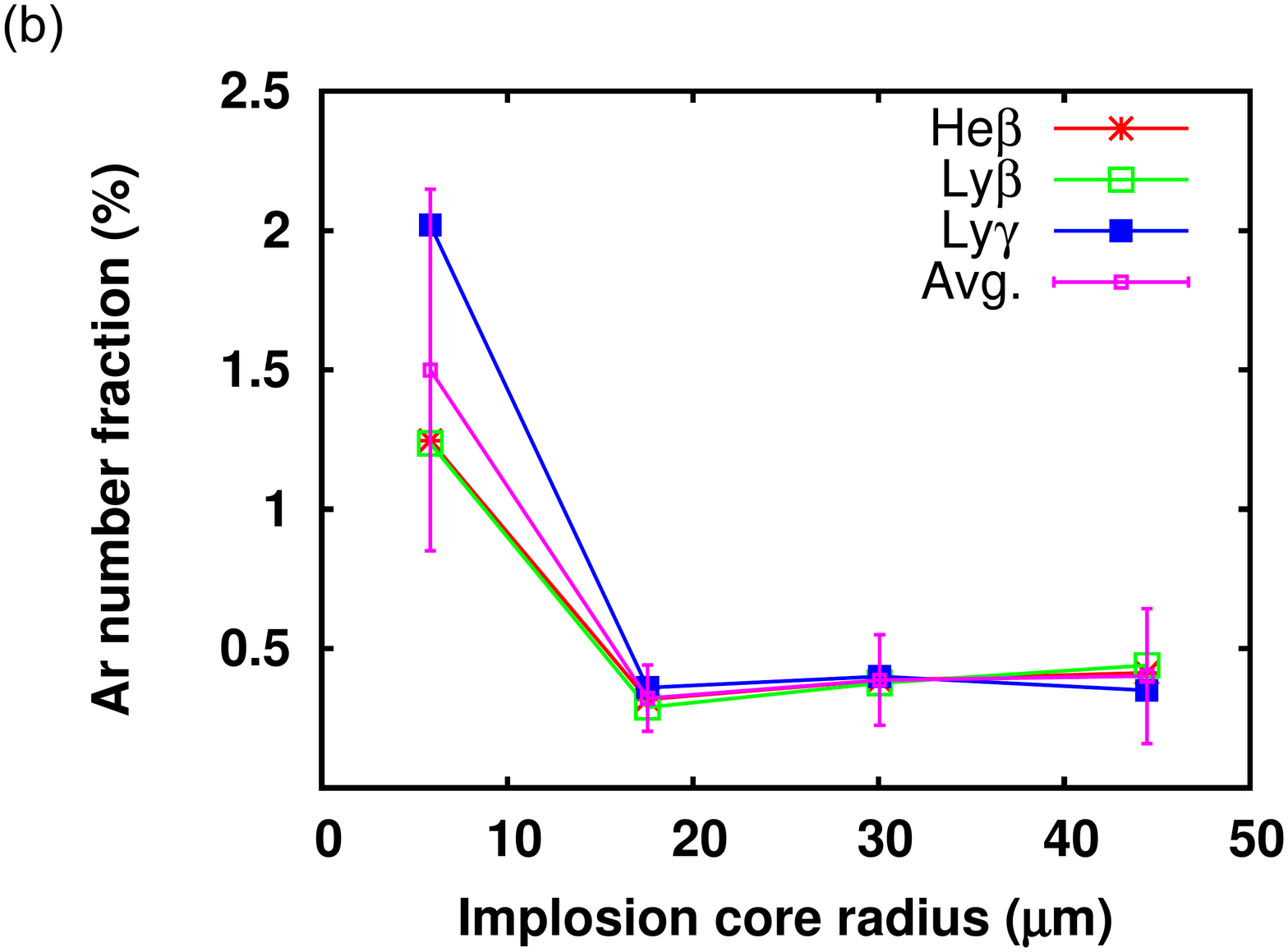}
\caption{(a) Deuterium atom number density, and (b) Ar atom number fraction spatial profiles extracted from the  He$\beta$, Ly$\beta$ and Ly$\gamma$ lines, respectively, in the implosion core of OMEGA shot 78199 TIM 3 Frame 2}
\end{figure}

\subsection{Symmetry analysis of narrow-band images}
In order to apply the generalized Abel inversion\cite{Welser-Sherrill07pre} in narrow-band images, the images should be axially symmetric. We observe some inhomogeneities in our narrow-band images shown in Fig.\:8a. To study the effect of these inhomogeneities on the $n_e$ and $T_e$ spatial profiles extracted from the images above, we divided He$\beta$ and Ly$\beta$ narrow-band images into three wedges of 120$^{\circ}$ each,  and each wedge was further divided into four radial zones. The number of wedges and radial zones were determined based on the spatial resolution of the MMI instrument. We then extracted intensity, emissivity, $n_e$ and $T_e$ spatial profiles for all three regions. Fig.\:8a shows the narrow-band images divided into three wedges of 120$^{\circ}$ each; 8b and 8c show spatial profiles of $T_e$ and $n_e$ extracted from all three wedges. All the profiles of $n_e$ and $T_e$ are consistent with the results obtained from the complete ``Annular" regions and the standard deviations across the three wedges are similar to the error bars shown in the Figs.\:5a and 5b. These results justify our assumption of 1D spherical symmetry in the analysis within the constraints of the spatial resolution of the instrument.

\begin{figure}[h!]
\centering
\includegraphics[width=8cm]{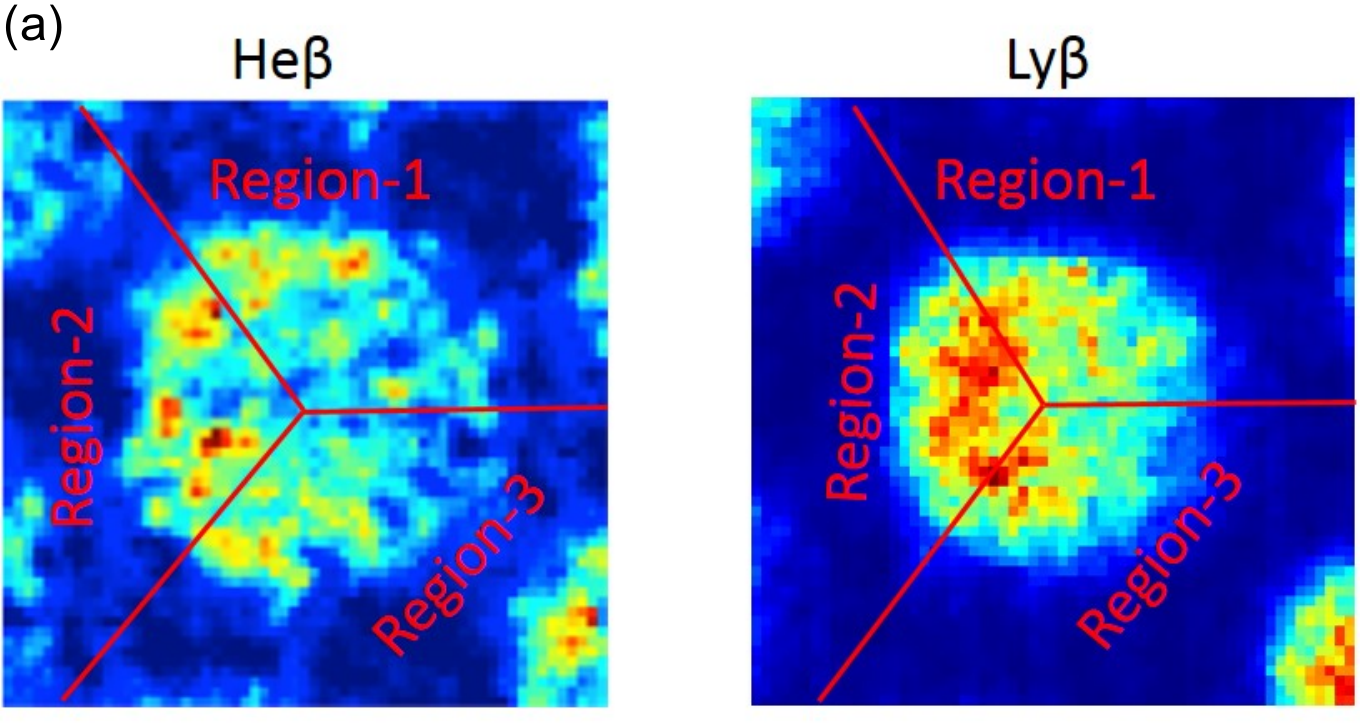}
\includegraphics[width=8cm,height=6.5cm]{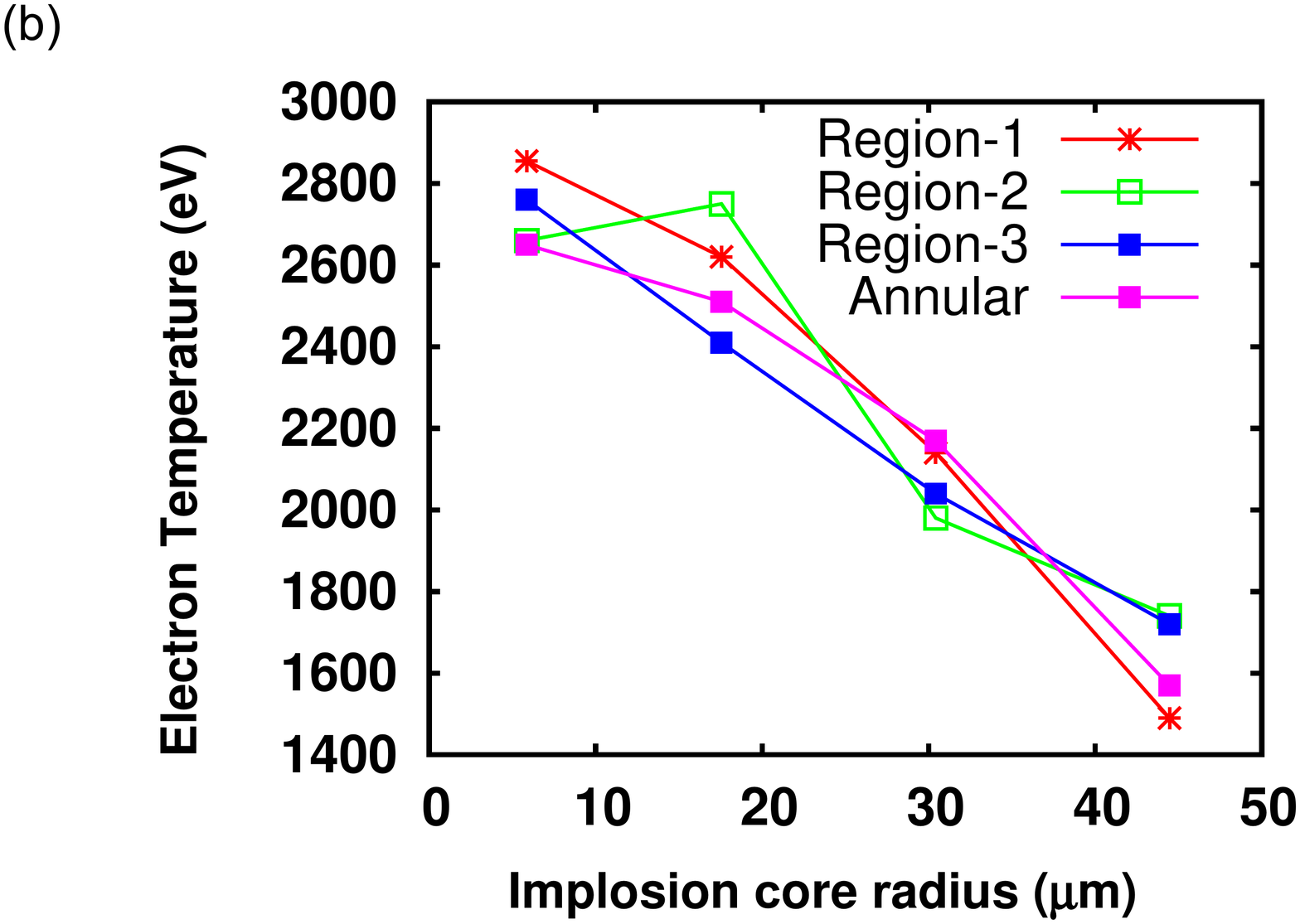}\\
\vspace{-0.5cm}
\includegraphics[width=9.1cm,height=6.5cm]{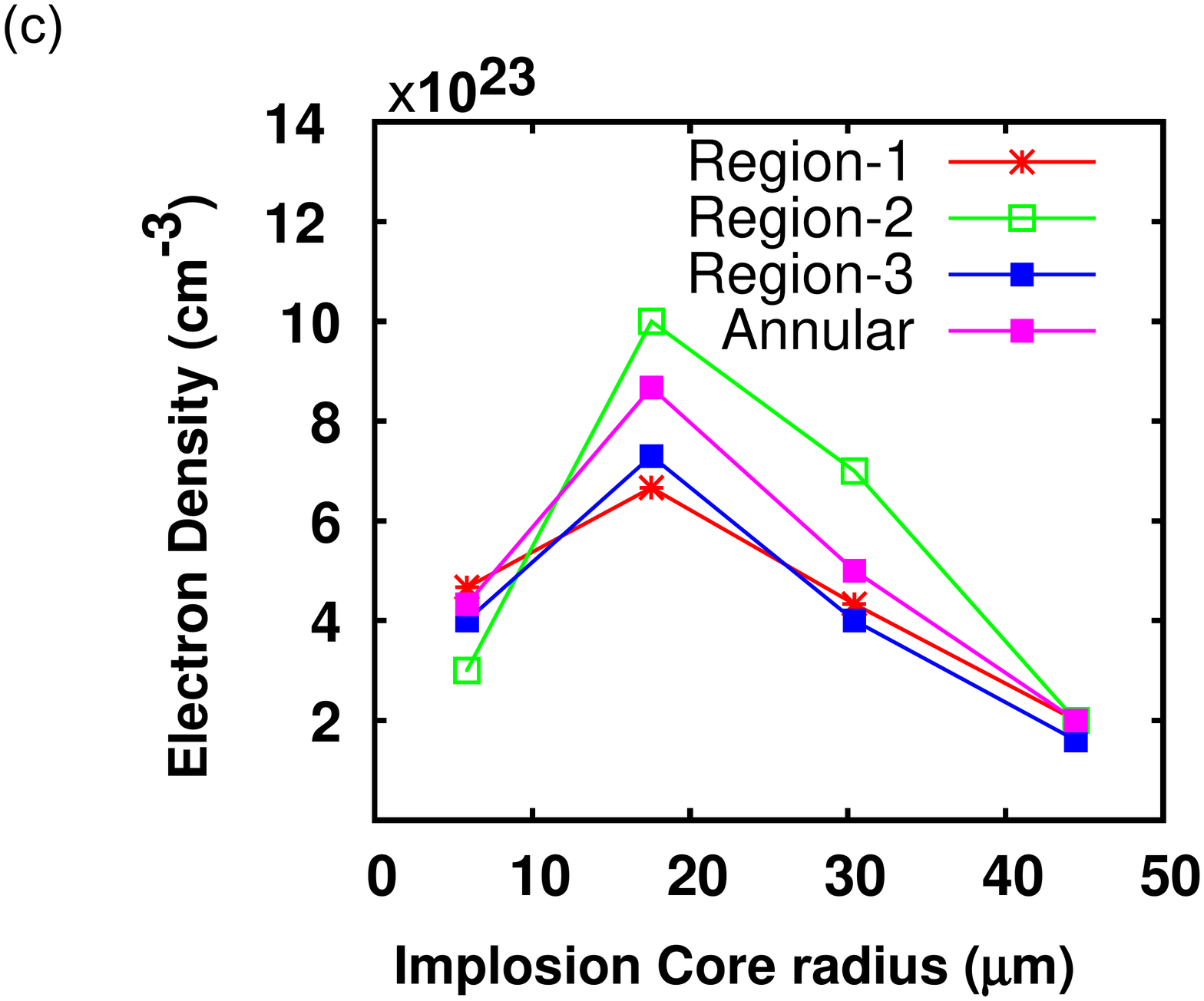}
\caption{(a) Narrow-band images divided into three wedges of 120$^{\circ}$. (b) Electron temperature and (c) density radial profiles extracted from all three wedges.}
\end{figure}

\subsection{Synthetic data analysis}
To accomplish the verification of our analysis method, we generated synthetic Ar He$\beta$, Ly$\beta$ and Ly$\gamma$ narrow-band images and space-resolved spectra using the same spectral range, $n_e$, and $T_e$ as in the case of OMEGA shot 78199 TIM 3 Frame 2 experimental results, and then used the same data analysis method to infer Ar atom number fraction vs. core radius. The synthetic data was generated with the FESTR code in forward reconstruction mode\cite{Hakel16cpc} and assuming 1D spherical symmetry. Figure\:9 displays results obtained from the analysis of synthetic narrow-band images and space-resolved spectra. Figure\:9a shows Ar atom number density ($n_{Ar}$) vs. core radius. Similarly, 9b  and 9c, respectively, show D atom number density ($n_D)$ and Ar atom number fraction vs. core radius. The trends of the results are similar and differences in the values are within the error included in the experimental data analysis. There is a little more spread in the central values in 9a and 9c; this could be due to the approximations of opacity  corrections (i.e., use of effective volumes) to our intensities (Eqs. (2a)-(2d)) and the uncertainties considered in our analysis method.

\begin{figure}
\centering
\includegraphics[width=8cm]{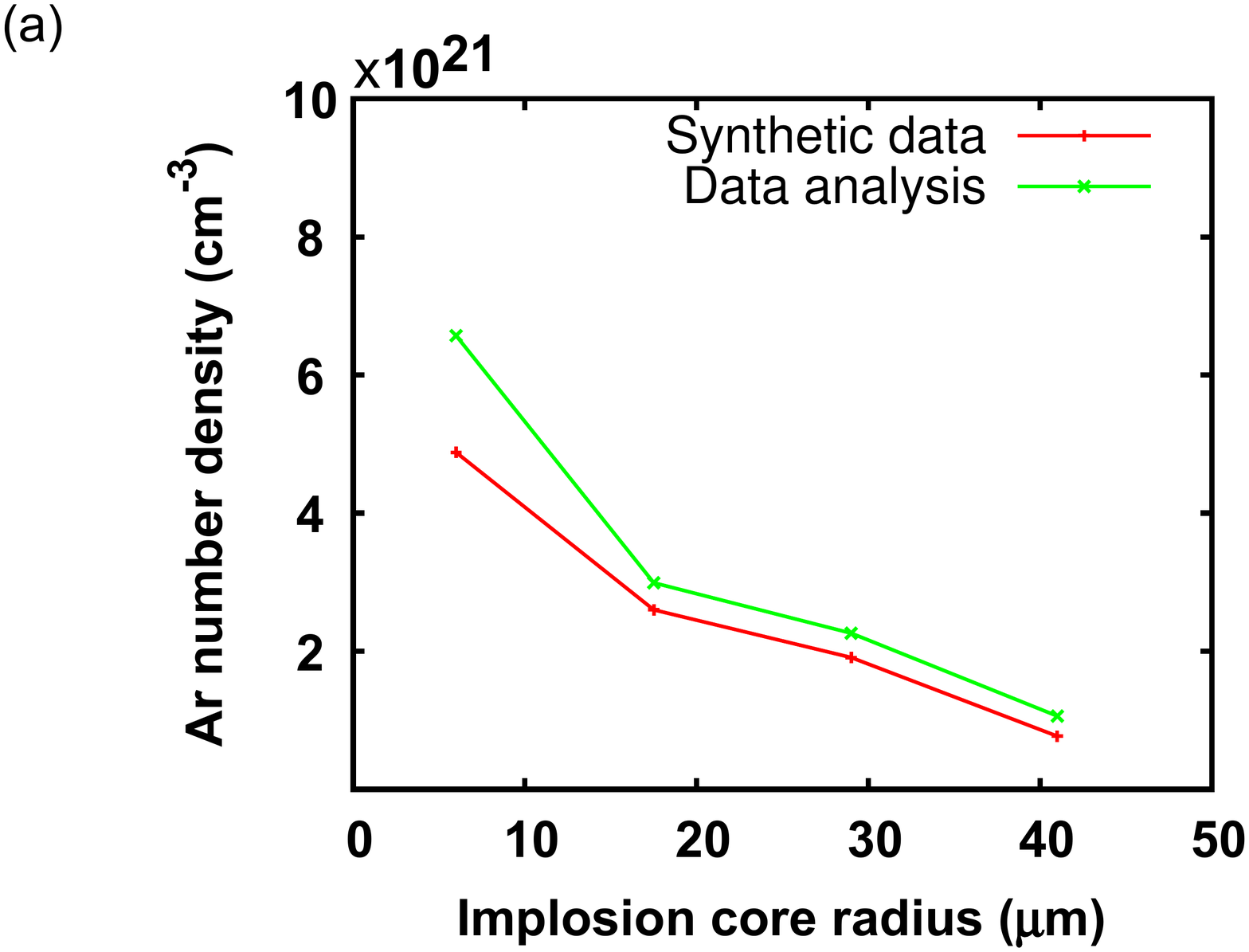}\\
\vspace{-1cm}
\includegraphics[width=8cm]{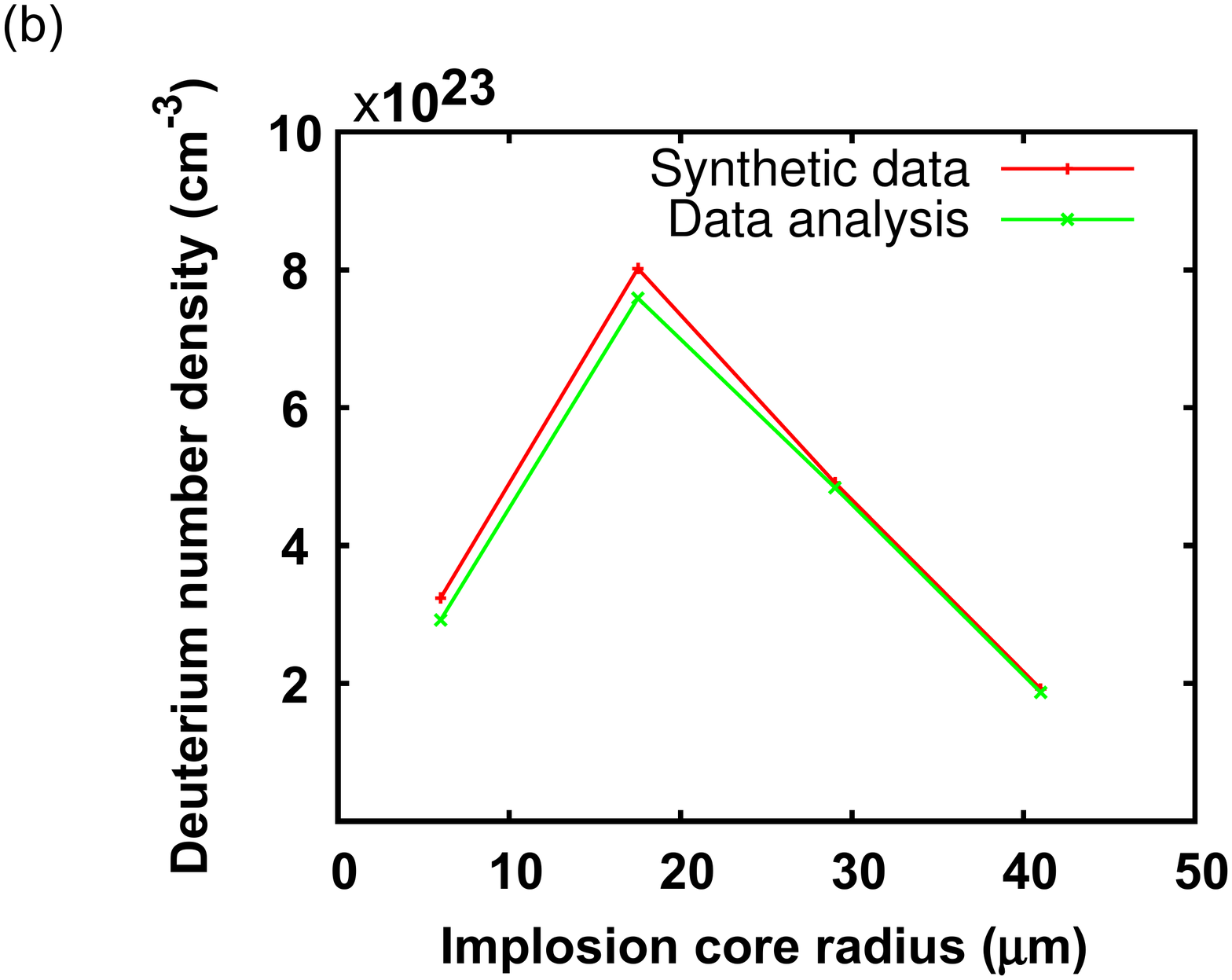}\\
\vspace{-1cm}
\includegraphics[width=8cm]{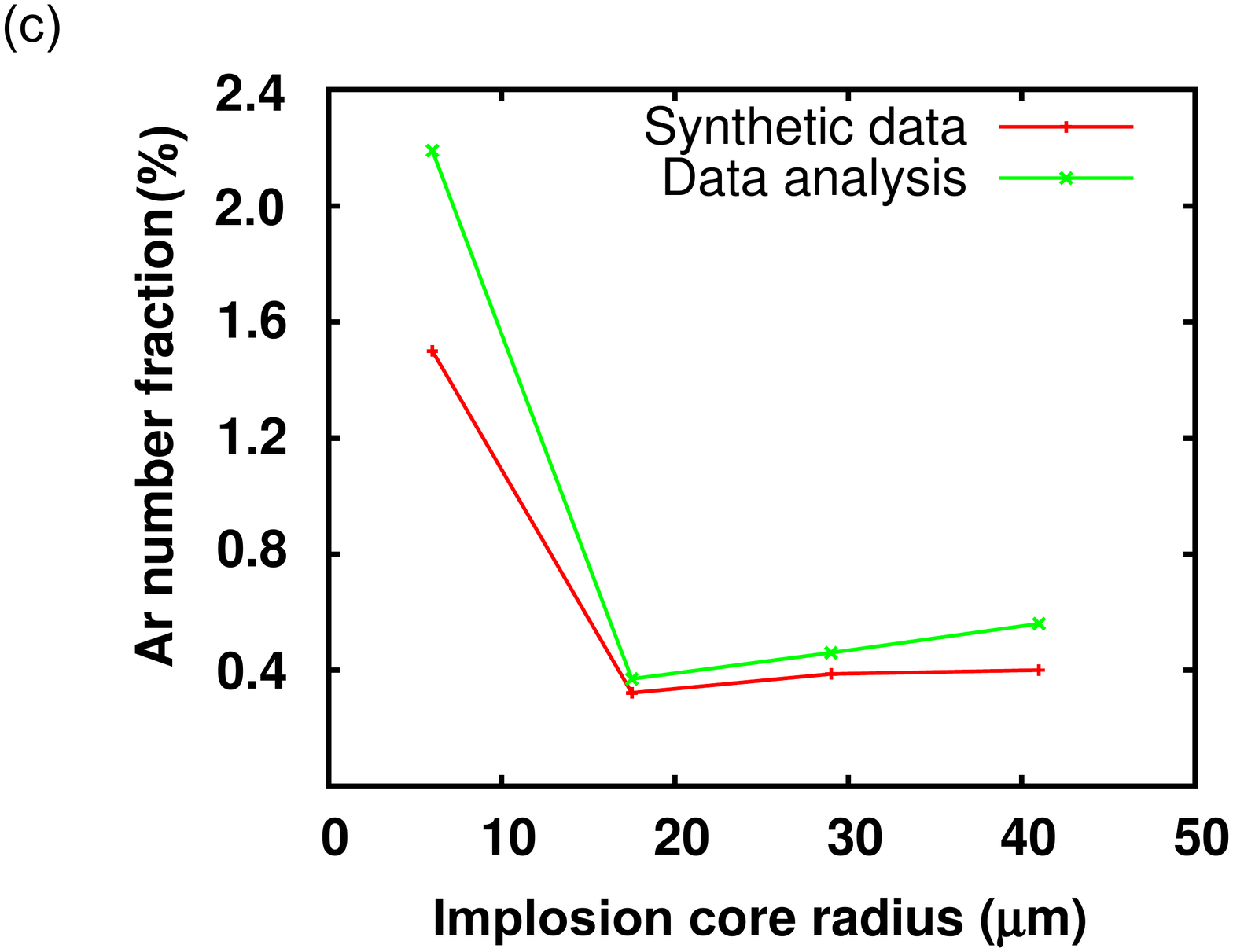}
\caption{Analysis of the synthetic narrow-band images and space-resolved spectra (equivalent to the case of OMEGA shot 78199 TIM 3 Frame 2). (a) Ar atom number density vs. implosion core radius, (b) deuterium atom number density vs. implosion core radius, and (c) Ar atom number fraction vs. implosion core radius.}
\end{figure}

\subsection{Error analysis in the experimental data}
The potential sources of uncertainty in experimentally inferred $T_e$ are: 1) the resolution of the electron temperature in the theoretical database (100 eV), 2) asymmetry in the narrow-band images, and 3) theoretical emissivity ratios of Ly$\beta$ to He$\beta$ as a function of electron temperature computed at constant $n_e$ with the assumption that the emissivity ratio depends  strongly on electron temperature but weakly on electron density. Figure\:10a shows the uncertainties in $T_e$ as a function of the core radius due to all the sources mentioned above. The red curve shows $T_e$ resolution in the database as a function of the core radius. The green curve shows asymmetry in $T_e$ (standard deviation in $T_e$) vs. core radius across the three wedges of the narrow-band images shown in the Fig.\:8a. To study the sensitivity of the ratio of the theoretical emissivity (Ly$\beta$/He$\beta$) vs. $T_e$ on $n_e$, we evaluated the ratio at four different $n_e$ values in our regime of interest (5.0$\times10^{23}$, 8.0$\times10^{23}$, 1.0$\times10^{24}$, 2.0$\times10^{24}$ cm$^{-3}$). The blue curve shows the standard deviation in $T_e$, obtained from the temperature spread due to the use of different emissivity ratio tables evaluated at electron densities mentioned above, as a function of core radius. The final error for the experimental $T_e$ in each spatial zone in the implosion core is obtained based on the greatest of the uncertainties in the above mentioned three curves. 

Similarly, Fig.\:10b shows uncertainties in the experimental $n_e$. The red curve shows resolution of $n_e$ in the theoretical database as a function of implosion core radius. The green curve shows asymmetry in $n_e$ due to the non-uniformities in the narrowband images shown in Fig.\:8a (standard deviation of $n_e$ across the three wedges of the narrowband images). Finally, the blue curve represents the standard deviation (SD) of $n_e$ obtained from the differences in the results from the three Ar K-shell lines present in the MMI data. Again, the final error for the experimental $n_e$ in each spatial zone in the implosion core is obtained based on the greatest of the uncertainties in the above mentioned three curves. Furthermore, we have observed that the error on $n_e$ due to the uncertainty in $T_e$ is either zero or equal to the resolution of $n_e$ in the database. The tabulation errors (temperature-density grid steps in the atomic database) can be reduced by performing 2D interpolation. For the cases considered here, the tabulation errors (see Fig.\:10) are approximately similar to the other errors for the most part, and do not alter the final conclusions of the error analyses.

\begin{figure}
\centering
\includegraphics[width=8cm,height=6.5cm]{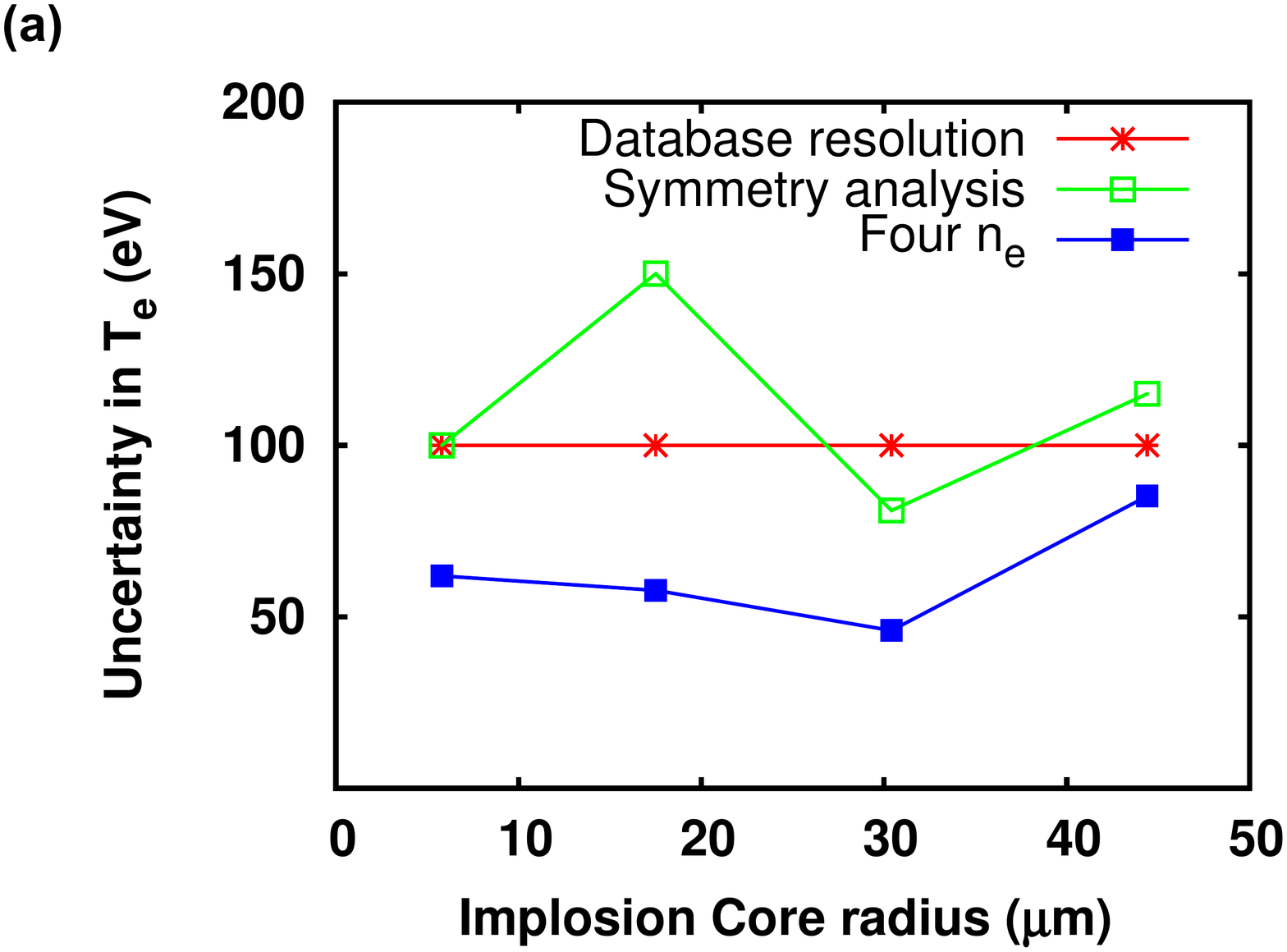}\\
\vspace{-0.5cm}
\includegraphics[width=8.9cm,height=6.5cm]{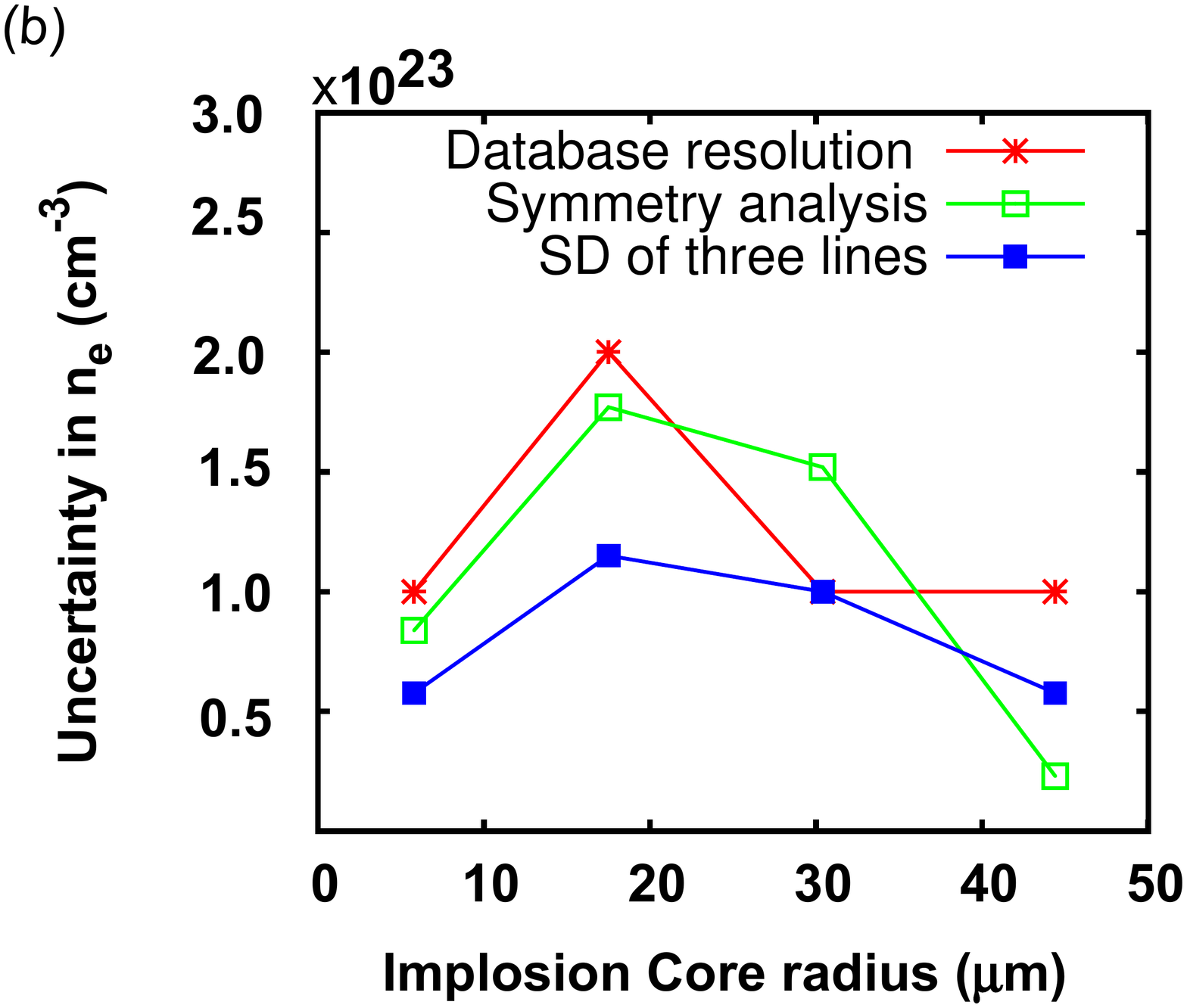}
\caption{Sources of uncertainty in the experimentally inferred $T_{e}$ and $n_e$ shown in Fig.s 5a and b, and 6b. (a) Uncertainties in $T_e$ vs. implosion core radius. (b) Uncertainties in $n_e$ vs. implosion core radius.}
\end{figure}

From Eq.\:(3), it can be seen that the total error in $n_{Ar}$ depends on the error propagation from $n_u$ and $F_u(T_e,n_e)$. The error in $n_u$ is determined based on the uncertainty associated with the photon-energy integrated intensity (total intensities in Eqs.\:(2a)-(2d)) because of the uncertainty in the subtraction of the continuum level. To find the error in $n_u$, we first extract the intensities of a line from each annular space-resolved spectrum several times by re-estimating continuum levels with a slightly different (slightly higher or lower with each other) spectral ranges. After plugging these intensities into Eqs.\:(2a)--(2d), we obtain a set of $n_u$ values for a line in each spatial zone in the implosion core. Then, we find standard deviations of $n_u$ values of a line in each spatial zone and use them as the final errors in the upper-level population densities of that line. 
The error in $F_u(T_e, n_e)$ depends on the error propagation from $T_e$ and $n_e$, and can be obtained by using Eq. (6) (based on Eq.\:(3.14) of Ref.\:46)
\begin{equation}
{\sigma_{F_{u}}}^2\simeq\left(\frac{\partial F_u}{\partial T_e} \right )^2_{n_{e}}{\sigma_{T_e}}^2+\left(\frac{\partial F_u}{\partial n_e} \right )^2_{T_{e}}{\sigma_{n_e}}^2,
\end{equation}
where the first term represents the error associated with the change of fractional upper-level population density with respect to the change of electron temperature at constant electron density, and the second term represents the error associated with the change of fractional upper level population with  respect to the change of electron density at constant electron temperature. The outermost zone shows a higher percentage error because of the higher percentage error in the electron density in that zone. Similarly, the error propagation formula for $n_{Ar}$ is given in Eq.\:(7a). Equation\:(7b) can be obtained by plugging Eq.\:(3) in Eq.\:(7a).The final error of $n_{Ar}$ was obtained by propagating errors on $n_u$ and $F_u$ to Eq.\:(7b),
\begin{table*}[tb]
\caption{\label{table:78199t3f2_summary}Summary of the results from our analysis for OMEGA shot 78199 TIM 3 Frame 2.}
\begin{ruledtabular}
\begin{tabular}{llllll}
Radial Zones \# &$T_e\pm\Delta T_e$ &  $n_e\pm\Delta n_e$&$n_{Ar}\pm\Delta n_{Ar}$& $n_D\pm\Delta n_D$ &$f_{Ar}\pm\Delta f_{Ar}$\\
 &(eV)&                                  $(10^{23}$cm$^{-3})$&               $(10^{21}$cm$^{-3})$&                             $(10^{23}$cm$^{-3})$&               ($\%$)\\                                                                                                                                                            
 \colrule
 a & 2650$\pm$100 & $4.3\pm1.0$ & $4.9\pm1.4$ & $3.2\pm1.1$ & 1.50$\pm$0.65 \\
 b & 2510$\pm$150 & $8.7\pm2.0$ & $2.6\pm0.7$ & $8.0\pm2.1$ & 0.32$\pm$0.12 \\
 c & 2170$\pm$100 & $5.0\pm1.6$ & $1.9\pm0.5$ & $4.9\pm1.5$ & 0.38$\pm$0.16\\
 d & 1570$\pm$115 & $2.0\pm1.0$ &$0.8\pm0.4$ & $1.9\pm1.0$ & 0.40$\pm$0.30\\
\end{tabular}
\end{ruledtabular}
\end{table*}

\begin{subequations}
\begin{align}
&\sigma_{n_{Ar}}^2\simeq\left(\frac{\partial n_{Ar}}{\partial n_u} \right)^2_{F_{u}}\sigma_{n_u}^2+\left(\frac{\partial n_{Ar}}{\partial F_{u}} \right )^2_{n_{u}}\sigma_{F_{u}}^2\\
&\sigma_{n_{Ar}}^2\simeq\left(\frac{\sigma_{n_{u}}}{n_{u}}\right)^2 n_{Ar}^2+\left(\frac{\sigma_{F_{u}}}{F_{u}}\right)^2 n_{Ar}^2.
\end{align}
\end{subequations}

 Equations.\:(8a) and (8b) below give the error propagation formula for $n_{D}$ (using the  ionization balances of Ar and D, i.e., $Z_{Ar}$ and $Z_{D}$ as constants, since changes in $Z_{Ar}$ and $Z_{D}$ due to the uncertainties in $T_e$ and $n_e$ are very small). Equation\:(8b) can be obtained by plugging the charge-neutrality condition given by Eq.\:(4) in Eq.\:(8a). We calculated errors in $n_{D}$ extracted from all the three lines.
\begin{subequations}
\begin{align}
&\sigma_{n_D}^2\simeq\left(\frac{\partial n_D}{\partial n_e} \right )^2_{n_{Ar}}\sigma_{n_e}^2+\left(\frac{\partial n_D}{\partial n_{Ar}} \right )^2_{n_{e}}\sigma_{n_{Ar}}^2\\
&\sigma_{n_D}^2\simeq\frac{1}{Z_{D}^2}\sigma_{n_e}^2+\frac{Z_{Ar}^2}{Z_{D}^2}\sigma_{n_{Ar}}^2
\end{align}
\end{subequations}
Finally, the error propagation formula for $f_{Ar}$ is given in Eqs.\:(9a) and (9b). The latter can be obtained by plugging the value of $f_{Ar}$ from Eq.\:(5) in Eq.\:(9a). We calculated error in $f_{Ar}$ by propagating errors on $n_{Ar}$ and $n_D$ to Eq.\:(9b). Table I summarizes all the results including errors from our analysis for OMEGA shot 78199 TIM 3 Frame 2.

\begin{subequations}
\begin{align}
&\sigma_{f_{Ar}}^2\simeq\left(\frac{\partial f_{Ar}}{\partial n_{Ar}}\right )^2_{n_{D}}\sigma_{n_{Ar}}^2+\left(\frac{\partial f_{Ar}}{\partial n_{D}} \right )^2_{n_{Ar}}\sigma_{n_{D}}^2\\
&\sigma_{f_{Ar}}^2\simeq\left(\frac{\sigma_{n_{Ar}}}{n_{Ar}}\right)^2 f_{Ar}^2+\left(\frac{\sigma_{n_{D}}}{n_{D}}\right)^2 f_{Ar}^2
\end{align}
\end{subequations}

\section{Comparison of MMI results with hydrodynamic simulations}

\begin{figure}
\centering
\includegraphics[width=8cm]{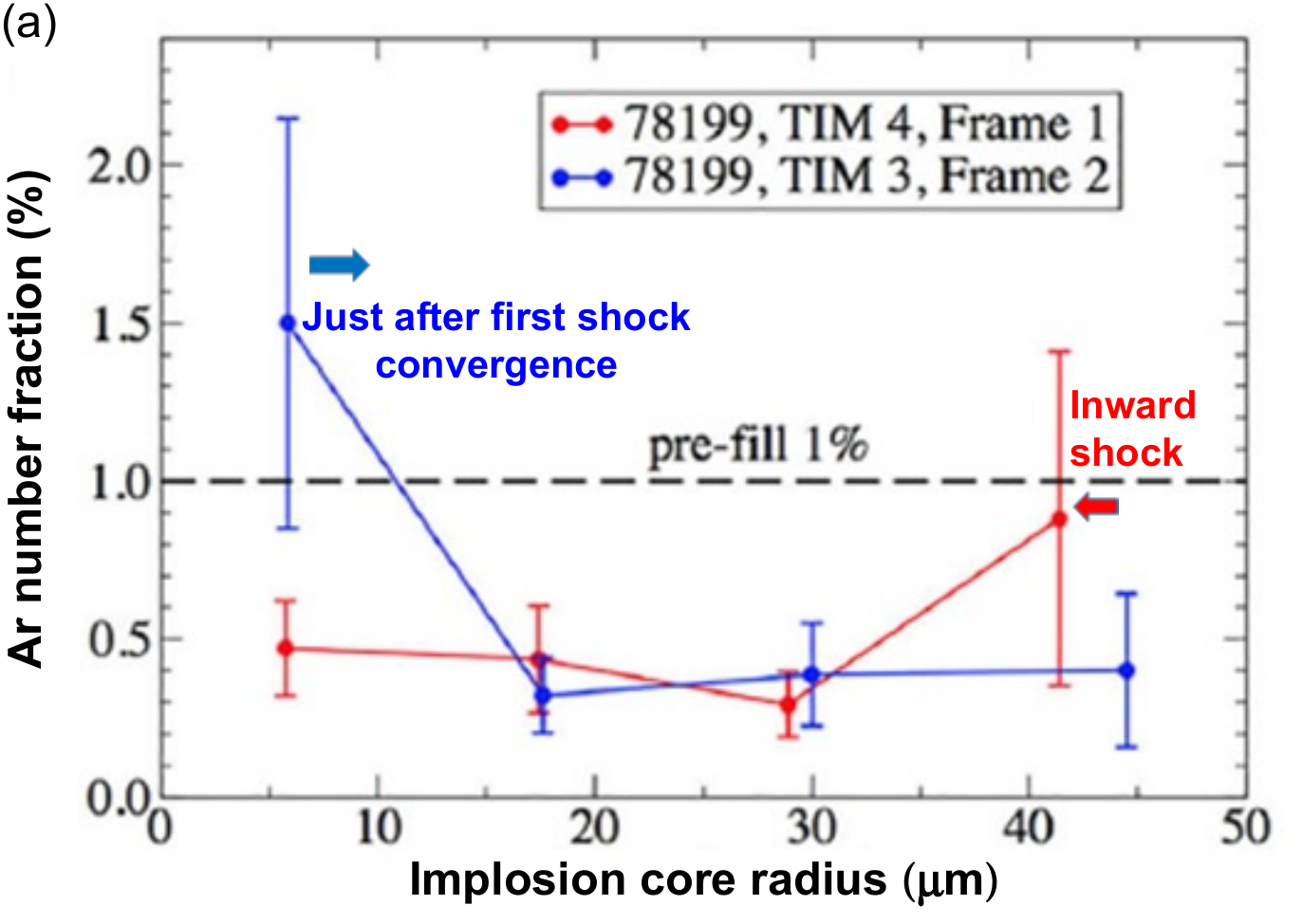}\\
\vspace{0.5cm}
\includegraphics[width=7.9cm]{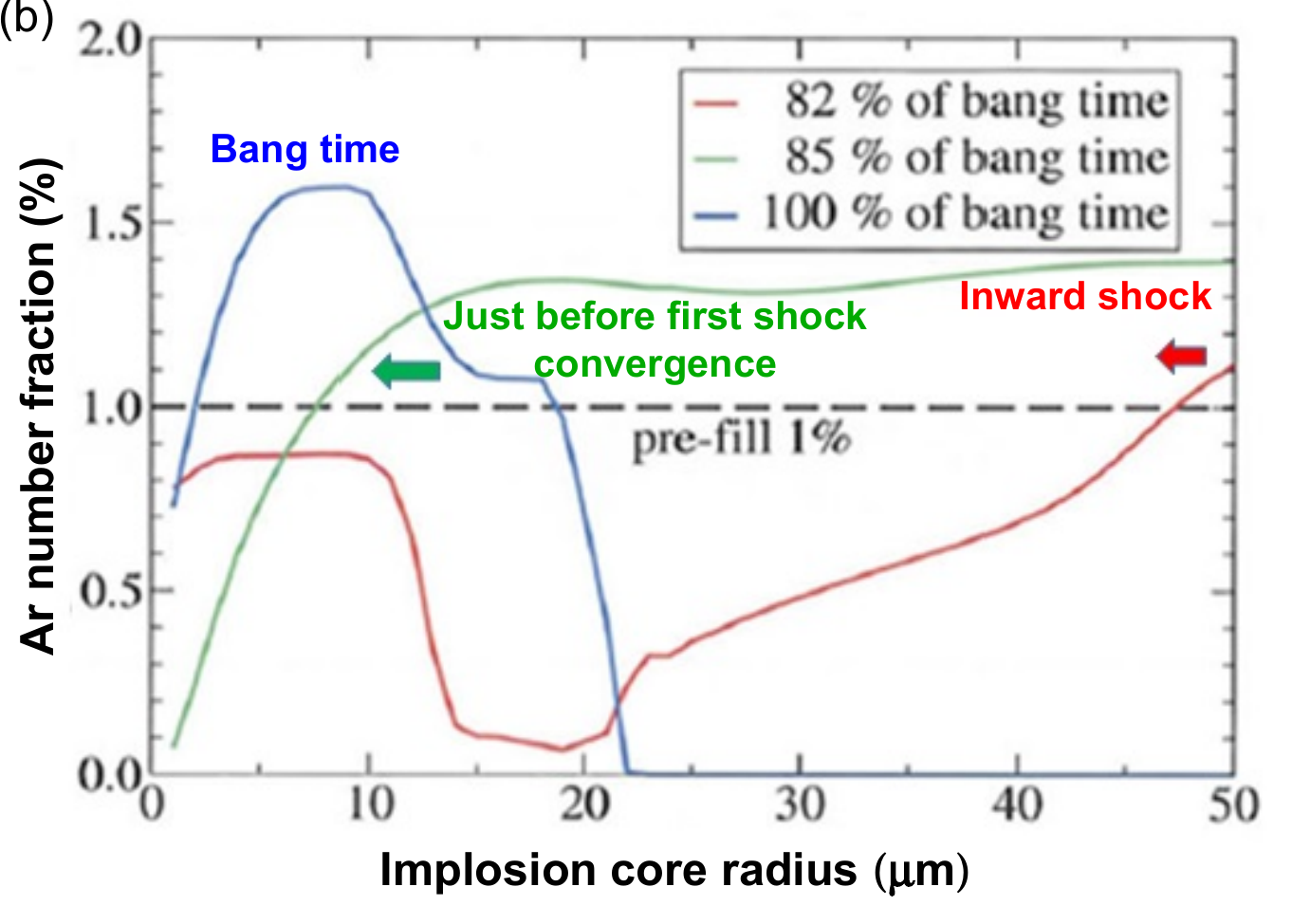}
\caption{(a) Ar atom number fraction  vs. implosion core radius for shot 78199. Red curve (84$\%$ of BT) is 40 ps later in time than blue (87$\%$ of BT) (b)  Ar atom number fraction  vs. implosion core radius obtained from the 1D post-shot xRAGE simulation of shot 78199 including a two-ion-species transport model.\cite{Gittings08csd, Molvig14pop}}
\end{figure}
Next, we performed 1D simulations of shot 78199 using xRAGE\cite{Gittings08csd} with a two-ion-species transport model\cite{Molvig14pop} to support our results obtained from the analysis method explained in the previous section. The brief comparisons between the experimental and simulated $f_{Ar}$ were previously reported.\cite{hsu_16_epl} We provide further comparisons and interpretations of the experimental results and simulations in this paper. Figure\:11a shows the experimentally inferred $f_{Ar}$ vs. $r$ (implosion core radius) for shot 78199 from the data recorded by two MMIs fielded along quasi-orthogonal directions, i.e., TIM 3 Frame 2 and TIM 4 Frame 1 with a trigger-time difference of approximately 40 ps. The bang time (BT) recorded by cryo-NTD for this shot was 1.313 ns, and trigger times for TIM 4 Frame 1 and TIM 3 Frame were, respectively, 1.10 ns (red curve, 84$\%$ of BT) and 1.14 ns (blue curve, 87$\%$ of BT). Figure\:11(b) shows $f_{Ar}$ for the same shot obtained from the 1D simulations mentioned above. The red, green and blue curves represent, respectively, spatial profiles of $f_{Ar}$ at 1.10 ns ($\approx$ 82$\%$ of BT), 1.15 ns ($\approx$ 85$\%$ of BT) and 1.35 ns ($\approx$ 100$\%$ of BT); all three curves show Ar depletion ahead of the incoming (or rebounding) shock followed by enhancement behind the incoming (or rebounding) shock. Furthermore, when we turned off the ion thermo-diffusion in the simulation, the strongest expected contributor to interspecies separation between D and Ar, in the calculation, no significant Ar-concentration variation from the 1$\%$ uniform pre-fill was obtained. The direction and location of the shocks are represented by the arrows in Figs.\:11a and 1b. 

While a detailed validation of the species-separation model is beyond the scope of this work, we have found resonable agreement between data and simulation of shot 78199. The magnitude of the peak in $f_{Ar}(\approx1.5\%$) agrees well between experiment (Fig.\:11a) and simulation (Fig.\:11b). The overall effects of incoming/rebounding shocks on species separation as predicted by the simulation, i.e., argon depletion (equivalently, D enhancement) in front of the incoming shock and argon enhancement behind the incoming and rebounding shocks, are also seen in the experimental results. Specifically, the red curve of Fig.\:11a at 1.10 ns shows lower argon concentration for r $\le$ 30 $\mu$m, which we interpret to be ahead of the incoming shock (presumed to be located around 40 $\mu$m at that time). However, there is less correspondence in timing between experiment and simulation. In the simulation (as shown in Fig.\:11b), shock flash occurs just after the time of the green curve ($\approx$85$\%$ of bang time in the simulation), at which time $f_{Ar}\ge$ $1\%$ for $r \ge$ 10 $\mu$m, suggesting that the incoming shock is located around 10 $\mu$m at that time. In the experimental result (Fig.\:11a), our interpretation is that the incoming shock at 1.10 ns (red curve at $\approx$ 84$\%$ of BT) is located around 40 $\mu$m. Forty picoseconds later (blue curve at 1.14 ns, 87$\%$ of BT) our interpretation is that the shock has just rebounded and hence $f_{Ar}$ $>$ 1$\%$ at the origin. Thus, there is qualitative agreement between experiment and simulation in the time evolution of the expected behavior of $f_{Ar}$ from earlier to later in time, even though there is not exact time correspondence in the absolute time relative to BT. One possible explanation for the key differences between experiment and simulations is that, in reality, there is more D enhancement ahead of the shock front due to kinetic streaming of the D ahead of the shock. This does not occur in the fluid-based model of our simulation. Secondly, our xRAGE simulations (at the moment) model binary interspecies diffusion (in our case, between D and Ar in the hot spot), and do not model the interspecies diffusion across the gas-shell interface. Thus the xRAGE predictions of $f_{Ar}$ are subject to the lack of CH mixed into the hot spot. In future work, we will be able to better assess this difference, as there are plans to include N-species ion diffusion in xRAGE, which would allow us to explicitly model shell/gas interspecies ion diffusion.

\section{Conclusions}
We have reported the details (experiments and analysis method) of the first direct experimental evidence for interspecies ion separation in direct-drive ICF experiments\cite{hsu_16_epl} performed at the OMEGA laser facility, via detailed analyses of imaging x-ray spectroscopy data. Our targets were spherical plastic shells filled with D$_2$ and Ar, designed to maximize interspecies ion thermo-diffusion between ion species of  large mass and charge difference. Ar K-shell spectral features were observed primarily between the time of first-shock convergence and slightly before the neutron bang time, using a time- and space-integrated spectrometer, streaked crystal spectrometer, and two gated multi-monochromatic x-ray imagers (MMI) fielded along quasi-orthogonal lines of sight. The spectrally resolved MMI data were processed to obtain narrow-band images and radially resolved spectra. Our analysis method to infer Ar atom number fraction spatial profiles used three steps. In the first step, we extracted the electron temperature and density spatial distribution by performing analyses of emissivity profiles of K-shell lines extracted from the MMI narrowband images.\cite{Welser-Sherrill07pre} In the second step, spatial distributions of Ar atom number density  were obtained by using a new inversion method\cite{Joshi15} based on the annular space-resolved spectra extracted from the MMI data. In the final step, we used the charge quasi-neutrality constraint to obtain deuterium number density and Ar atom number fraction $f_{Ar}$. The details of the error propagation techniques and quantitative values for shot 78199 TIM 3 Frame 2 have been discussed. Errors in the final results were obtained by propagating uncertainties from step 1 to step 3 including all the possible sources of uncertainties in each step. The simulations show that $f_{Ar}$ is enhanced and depleted, respectively, in front of and behind the first incoming shock and our data agree reasonably well with the calculated effects on $f_{Ar}$ due to the incoming and rebounding first shock.\cite{hsu_16_epl}

In forthcoming work, we would like to quantify the effect of shell/gas mix on our extraction of $f_{Ar}$ by using $N$-species modeling and extracting CH concentration in the core. Similarly, we would like to quantify Ar diffusion into the shell and its effect on our extraction of $f_{Ar}$. Also, we want to obtain argon density by absolute photo-calibration of MMI data, and compare with the present result that assumes conservation of argon atoms. We were unable to observe the experimental Ar K-shell spectral signatures late in time in the implosion (around the time of the peak compression), and thus are unable to report whether interspecies ion separation is observed through bang time. In future experiments we could try a higher-$Z$ element other than Ar in some of our targets to observe whether species separation persists at bang time (we believe that Ar may be completely ionized at the temperature and density conditions achieved at peak compression/neutron bang time). Nevertheless, this first result using x-ray spectroscopy to detect interspecies ion separation encourages further work toward establishing a validated capability to address its role in yield degradation of ignition-scale ICF implosions.\cite{hsu_16_epl} Future work will also include use of Pareto genetic algorithm and Levenberg-Marquardt methods\cite{Nagayama12pop} with FESTR\cite{Hakel16cpc} to build an automated search and optimization technique to extract 3D profiles of tracer concentration in the implosion core.
\begin{acknowledgments}
We acknowledge R. Aragonez, T. Archuleta, J. Cobble, J. Fooks, V. Glebov, M. Schoff, T. Sedillo, C. Sorce, R. Staerker, 
N. Whiting, B. Yaakobi, and the OMEGA operations team for their support in experimental planning, execution,
and providing processed x-ray and neutron data.  This work was supported by the LANL ICF 
and ASC (Advanced Simulation and Computing) Programs under US-DoE contract no.\ DE-AC52-06NA25396.
\end{acknowledgments}

%

\end{document}